\documentclass[10pt,a4paper,english]{amsart} 

\usepackage[margin=3cm]{geometry}
\usepackage[T1]{fontenc}
\usepackage[utf8]{inputenc}
\usepackage{babel}
\usepackage{multicol}
\usepackage{amsthm,amssymb,amsmath}
\usepackage{bm}
\usepackage{bbm}
\usepackage{graphicx}
\usepackage{subfigure}
\usepackage{tikz}
\usepackage{array}
\usepackage{mathptmx}
\usepackage{multirow, makecell}
\usetikzlibrary{arrows,positioning,shapes.geometric}
\usetikzlibrary{matrix,chains,decorations.pathreplacing}

\usepackage{xcolor}
\usepackage{microtype}
\usepackage{booktabs}
\usepackage{url,diagbox}

\begin{document} 

\title[Sleep-wake classification via quantifying HRV by CNN]{Sleep-wake classification via quantifying heart rate variability by convolutional neural network}

\author{John Malik, Yu-Lun Lo, Hau-tieng Wu}

%\address{$^1$ Department of Mathematics, Duke University, Durham, NC, USA} 
%\address{$^2$ Department of Thoracic Medicine, Chang Gung Memorial Hospital, Chang Gung University, School of Medicine, Taipei, Taiwan} 
%\address{$^3$ Department of Statistical Science, Duke University, Durham, NC, USA}
%\address{$^\dagger$ \texttt{john.malik@duke.edu}}

\maketitle

%\noindent{\it Keywords\/}: convolutional neural network, instantaneous heart rate, heart rate variability, sleep stage

\begin{abstract}
Fluctuations in heart rate are intimately tied to changes in the physiological state of the organism. We examine and exploit this relationship by classifying a human subject's wake/sleep status using his instantaneous heart rate (IHR) series. 
We use a convolutional neural network (CNN) to build features from the IHR series extracted from a whole-night electrocardiogram (ECG) and predict every $30$ seconds whether the subject is awake or asleep. 
Our training database consists of $56$ normal subjects, and we consider three different databases for validation; one is private, and two are public with different races and apnea severities.
On our private database of $27$ subjects, our accuracy, sensitivity, specificity, and $\mathrm{AUC}$ values for predicting the wake stage are $83.1 \%$, $52.4\%$, $89.4\%$, and $0.83$, respectively. Validation performance is similar on our two public databases. When we use the photoplethysmography instead of the ECG to obtain the IHR series, the performance is also comparable. A robustness check is carried out to confirm the obtained performance statistics. 
This result advocates for an effective and scalable method for recognizing changes in physiological state using non-invasive heart rate monitoring. The CNN model adaptively quantifies IHR fluctuation as well as its location in time and is suitable for differentiating between the wake and sleep stages.
\end{abstract}

\section{Introduction}\label{Sect:Introduction}

Sleep is a key ingredient to our well-being, and disrupted sleep may lead to catastrophes in personal medicine or public health \cite{Karni1994,Kang2009,Colten_Altevogt:2006}. Understanding sleep is critical for the whole healthcare system. The gold standard for quantifying sleep dynamics is the polysomnography (PSG). However, it is bulky, complicated to install, labor-intensive, and expensive. It is also possibly error-prone if an overnight PSG recording is manually examined by experts \cite{norman2000interobserver}. An automatic, convenient, and accurate way to study sleep is therefore in high demand. 
Efforts made in the past few decades to accomplish these tasks have been assisted by advancements such as bio-sensors, wearable technology, and mobile health devices. 

The assessment of sleep stage using vital signs is most commonly done using the electroencephalogram (EEG).  The photoplethysmography (PPG) and the electrocardiogram (ECG) are also widely considered.  Compared to a PSG, these signals are easy to install and cheap to obtain, and long-term monitoring is possible.  The challenge with these signals is that the amount of available information is limited.  Neural networks lend themselves advantageously to this task, as they excel at extracting and compressing information into representations that are useful for the application at hand \cite{LeCun2015}.
Motivated by the needs and challenges associated with predicting sleep dynamics using mobile health devices, we focus on analyzing ECG and PPG signals in this work, and we propose an effective and scalable approach to analyzing heart rate variability (HRV) as it pertains to physiological events and outcomes \cite{Malik_Camm:1995,Electrophysiology1043,Bravi2011}.

\subsection{Heart Rate Variability and Sleep}

HRV is quantified by studying the time series consisting of intervals between consecutive pairs of heart beats, which is usually called the R-to-R interval (RRI) time series. 
Interest in HRV has a long history \cite{Billman2011}, and it has been extensively studied in the past few decades. See, for example, a far-from-complete list of review articles \cite{Vanderlei_Pastre_Hoshi_Carvalho_Godoy:2009,Shaffer2014}. 
The rhythm of the heart is regulated by the autonomic nervous systems (ANS). Changes in heart rate are the result of complicated interactions between physiological systems and external stimuli.  They reflect the integrity of the physiological systems. In short, analyzing HRV amounts to observing through a non-invasive window the physiological dynamics of an organism. 
It has long been believed that a correct quantification of HRV will yield a deeper understanding of a variety of physiological systems, including sleep \cite{Snyder:1964,Zemaityte1984,Somers:1993,Vaughn1995,Toscani1996,Bonnet1997,SigridElsenbruchMichaelJHarnish1999,Chouchou2014,Penzel2016}. For example, while a subject is awake, the sympathetic tone of the ANS is dominant. The heart rate is higher due to daily activity and external stimuli, and the rhythm of the heart is less stable \cite{Somers:1993}.   
The heart rate is relatively lower when the subject is asleep. It reaches its lowest value during slow wave (deep) sleep \cite{Snyder:1964}. During non-REM (rapid eye movement) sleep, the parasympathetic nervous system dominates, the sympathetic tone becomes less dominant, and the energy restoration and metabolic rates reach their lowest levels. The heart rate decreases, and the rhythm of the heart stabilizes \cite{Somers:1993}. These physiological facts indicate that we can distinguish between the wake and sleep stages by taking heart rate into account. There have been several works in this direction, such as \cite{Lewicke2008,Mendez2010,Long2012,Xiao2013,Aktaruzzaman2015,Ye2016,Vicente2016}.   

With appropriate HRV quantities, physicians can better understand the ANS and hence improve both diagnostic accuracy and treatment quality \cite{Vanderlei_Pastre_Hoshi_Carvalho_Godoy:2009}.
However, despite a lot of research in this direction and many quantitative indices proposed by experts, there is limited consensus in the universality of HRV analysis. 
The reason for the lack of extensive success in this field is complicated due to the non-stationarity nature of the system. See, for example, some published discussions on this topic \cite{Pincus1994,Glass2009}. We do not expect to sort out this difficult problem in this exploratory work which focuses on the sleep stage classification problem.
Instead, we propose the use of a convolutional neural network (CNN) on the IHR series as a scalable approach to adaptively quantifying heart rate fluctuation from a variety of monitoring sources. A CNN \cite{LeCun1998} is a special kind of neural network (NN) capable of building features from time series and images. A CNN detects not only the presence of structured features which are important for classification, but also their temporal (spatial) location. CNN design is motivated by the hypothesized mechanism for the animal visual cortex. An important property of the CNN is that it behaves equivariantly with respect to translations of input features. This design is favorable for sleep stage classification because the precise locations of fluctuations which correspond to sympathetic (``fight or flight'') tone dominance provide physiological information about sleep stage. We refer readers with interest in CNN architecture to the review article \cite{LeCun2015}.   
There have been several studies using NNs and manually designed heart rate features to predict sleep stages \cite{Lewicke2005,Aktaruzzaman2015}, but to the best of our knowledge, CNNs have not been considered in the field.   
We show that the CNN approach is effective for accomplishing this task, providing evidence that it may lend itself favorably to other HRV applications.

\section{Materials and Methods}\label{sec: Materials}

\subsection{Data}

Standard overnight PSG studies were performed to confirm the presence of sleep apnea syndrome in clinical subjects suspected of sleep apnea at the sleep center in Chang Gung Memorial Hospital (CGMH), Linkou, Taoyuan, Taiwan. The Institutional Review Board of CGMH approved the study protocol (No. 101-4968A3). All recordings were acquired on the Alice 5 data acquisition system (Philips Respironics, Murrysville, PA). 
The sleep/wake stages were defined and scored by two experienced sleep technologists abiding by the AASM 2007 guidelines \cite{Iber_Ancoli-Isreal_Chesson_Quan:2007}, and a consensus was reached.  
Each recording is at least 5 hours long. We focus only on the PPG recording and on the second lead of the ECG recording, which were both sampled at $200$ Hz. There are 90 healthy subjects (each with apnea-hypopnea index less than $5$) in the training database, among which we consider only $56$ subjects who were labeled as awake for at least $10\%$ of the recording duration.  This database is called CGMH-training. 

We consider three validation databases. The validation databases were not used to tune the model's parameters, and were not subjected to any rejection. The first one consists of $27$ subjects and was acquired independently of CGMH-training from the same sleep laboratory. This database is called CGMH-validation.
The other two databases are publicly available. The DREAMS Subjects Database\footnote{\url{http://www.tcts.fpms.ac.be/~devuyst/Databases/DatabaseSubjects}}, consists of $20$ recordings from healthy subjects. The recordings were selected by the TCTS Lab to be of high clarity and to contain few artifacts.  The technologists used the Brainnet\texttrademark{} system (Medatec, Brussels, Belgium) to obtain the ECG recordings. The sampling rate is $200$ $\mathrm{Hz}$, and the minimum recording duration is $7$ hours. Although the race information is not provided, we may assume that since the database is collected from Belgium, its population constitution is different from that of the CGMH databases.
The third database is the St. Vincent's University Hospital/University College Dublin Sleep Apnea Database (UCDSADB) that is publicly available at Physionet \cite{Goldberger_Amaral_Glass_Hausdorff_Ivanov_Mark_Mietus_Moody_Peng_Stanley:2000}\footnote{\url{https://physionet.org/pn3/ucddb}} and which consists of $25$ subjects with sleep apnea of various severities. The technologists used Holter monitors to obtain the ECG recordings. The sampling rate is $128$ $\mathrm{Hz}$, and the minimum recording duration is $6$ hours.  We focus on the first ECG lead in this study. The DREAMS Subjects Database is chosen to assess the model's performance on recordings which come from subjects of a different race, and the UCDSADB is chosen to assess the model's performance on recordings which come from subjects with sleep disorders.

\subsection{Instantaneous Heart Rate}

We apply a standard automatic R peak detection algorithm to each ECG recording, which is a modification of that found in \cite{elgendi:qrs}. Let $\{r_i\}_{i=1}^n$ denote the location in time ($\mathrm{sec}$) of the $n$ detected R peaks. 
We ensure that there is no artifact in the detected R peaks by ignoring beats which are either too close or too far from their preceding beats. (The mechanism for detecting false positive or missing beats is based on a $5$-beat median filter.) We view the instantaneous heart rate (IHR) as a continuous and positive function.  We estimate the IHR at time $r_i$ in beats-per-minute ($\mathrm{bpm}$) as 
\begin{gather}
\mathrm{IHR}\left( r_i \right) = \frac{60}{r_i - r_{i-1}} \quad i = 2,...,n.
\end{gather}
Using these estimates, we follow the Task Force standard \cite{Electrophysiology1043} and obtain the IHR series over the duration of the recording at a sampling rate of $4$ Hz. The interpolation method is shape-preserving piecewise cubic interpolation.
When the PPG signal is considered, the same procedure is applied: we view the peaks in each PPG recording as surrogate heart beats, and these peaks are automatically detected by \cite{Elgendi2013ppg}. With the same interpolation scheme, the resulting time series is called IHR-PPG (for the sake of distinguishing it from the IHR series extracted from the ECG signal).
We break the IHR (or IHR-PPG) signal into $30$-second epochs, abiding by the boundaries originally assigned by the sleep technologists. We discard an epoch if fewer than five R peaks are detected within it. 

\subsection{Convolutional Neural Network}

We implement the $1$-dimensional CNN using TensorFlow 1.5 \cite{tensorflow2015-whitepaper}.
The feed-forward network architecture is shown in Figure~\ref{Fig:Network}. 
The input signal is first passed through five convolution blocks. Following common practice in HRV analysis \cite{Engoren:1998,Casaseca-de-la-Higuera_Martin-Fernandez_Alberola-Lopez:2006}, we consider input signals 5 minutes in length.  (Building an input signal $5$ minutes in length amounts to concatenating the labeled $30$-second epoch with the preceding $4$ minutes and $30$ seconds.) We normalize each $5$-minute input signal by subtracting its median value.

The architecture of a single convolution block is shown in Figure~\ref{Fig:ConvBlock}. Each convolutional layer in the block has $10$ filters with kernel size $8$.  A bias is added to the output of each filter, and the result is passed through a rectified linear unit (ReLU) activation function. The convolution blocks are followed by two fully-connected (dense) layers of $20$ nodes each. The fully-connected nodes also have associated biases and ReLU activation functions. We apply dropout with probability $0.5$ to the last convolutional layer and to both fully-connected layers. The output of the last fully-connected layer is fed with biases into a 2-node output layer.  We normalize the output of the network using the softmax function. We predict that a subject is awake during a given epoch if the output of the ``wake'' node is greater than or equal to the output of the ``sleep'' node. We train the network for $180$ epochs using mini-batch gradient descent with a learning rate of $10^{-3}$, a batch size of $24$, and cross-entropy as the loss function.

\tikzstyle{line} = [draw, -latex']
\tikzstyle{arrow} = [thick,->,>=stealth]

\begin{figure}
\centering
   \begin{tikzpicture}[>=latex']
        \tikzset{
        block/.style= {draw, rectangle, align=center,minimum width=5cm,minimum height=.10cm,line width=0.3mm},
        rblock/.style={draw, shape=rectangle,rounded corners=1.5em,align=center,minimum width=12cm,minimum height=.10cm},
        }

        \node [block]  (Input) {Input};
        \node [block, below =.3cm of Input] (Conv1) {Convolution Blocks};
        \node [block, below  =.3cm of Conv1] (Dense1) {$20$-Dense};
        \node [block, below  =.3cm of Dense1] (Dense2) {$20$-Dense};
        \node [block, below  =.3cm of Dense2] (Output) {Output};

          \draw[arrow]   (Input.south)  --(Conv1.north) node [pos=0.66,right] {};
          \draw[arrow]   (Conv1.south)  --(Dense1.north) node [pos=0.66,right] {};
          \draw[arrow]   (Dense1.south)  --(Dense2.north) node [pos=0.66,right] {};
          \draw[arrow]   (Dense2.south)  -- (Output.north) node [pos=0.66,right] {};
         
    \end{tikzpicture}
    \caption{\label{Fig:Network}The architecture of the $1$-dimensional convolutional neural network.  The notation $m$-Dense means that the fully connected layer possesses $m$ nodes.  For input signals $5$ minutes in length, we use five convolution blocks.}
\end{figure}
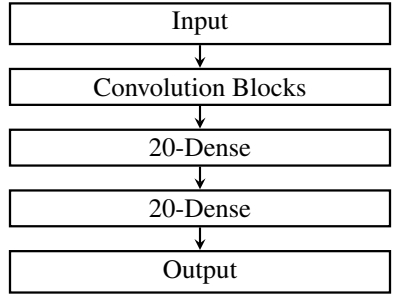

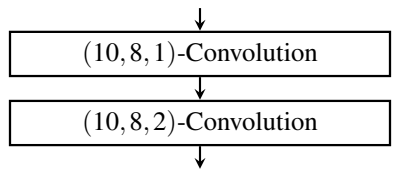
\begin{figure}
\centering
   \begin{tikzpicture}[>=latex']
        \tikzset{
        block/.style= {draw, rectangle, align=center,minimum width=5cm,minimum height=.10cm,line width=0.3mm},
        rblock/.style={draw, shape=rectangle,rounded corners=1.5em,align=center,minimum width=12cm,minimum height=.10cm},
        }

        \node (Input) {};
        \node [block, below  =.3cm of Input] (Conv1) {$(10, 8, 1)$-Convolution};
        \node [block, below  =.3cm of Conv1] (Dense) {$(10, 8, 2)$-Convolution};
        \node [below  =.3cm of Dense] (Output) {};

          \draw[arrow]   (Input.south)  --(Conv1.north) node [pos=0.66, right] {};
          \draw[arrow]   (Conv1.south)  --(Dense.north) node [pos=0.66, right] {};
          \draw[arrow]   (Dense.south)  -- (Output.north) node [pos=0.66,right] {};
         
    \end{tikzpicture}
    \caption{\label{Fig:ConvBlock}The architecture of a single convolution block.  The notation $(f, k, s)$-convolution means that the convolutional layer has $f$ filters with kernel size $k$ and stride $s$.  The output of the block is half the size of the input.}
\end{figure}

\subsection{Performance Evaluation}

We train the CNN model to differentiate between the wake and sleep stages by fitting the model's parameters to the IHR series extracted from CGMH-training. We then apply the trained model to three different validation databases: CGMH-validation, the DREAMS Subjects Database, and the UCDSADB. We report a variety of performance measures, namely sensitivity (recall), specificity, accuracy, AUC (area under the receiver operating characteristic curve), precision (PPV),  the F$1$ score, and Cohen's kappa coefficient.  We emphasize that our model assessment is performed on an inter-individual basis. 
We further access how effectively a model trained on signals from one monitoring device might be applied to recordings made on another device. The model trained on IHR series is applied to IHR-PPG series, and vice versa. (The PPG signal is commonly installed in modern mobile devices.) We compare the PPG performance statistics with those obtained using the corresponding ECG recordings.

\subsection{Robustness Check}

To check the robustness of the considered CNN model, we perform two sensitivity tests.  First, we consider different input sizes, namely $30$ seconds, $2$ minutes, and $10$ minutes. We assign a different number of convolution blocks to each input size. When the input size is 30 seconds (2 minutes and 10 minutes, respectively), we build up 3 (4 and 6, respectively) convolution blocks.  
In Figure~\ref{Fig:EpochLength}, we show that inputs of various sizes are constructed by additionally considering a length of time before the labeled epoch. 
In our second sensitivity test, we train new CNN models on the DREAMS Subjects Database and the UCDSADB. We validate these two models on our three validation databases. 
Note that the sampling rate for the UCDSADB is unique, and that the data collection system is heterogeneous across all databases; these observations allow us to test the robustness of the established CNN model.

\section{Results}\label{Section:Results}

Our training database from CGMH consists of recordings from $30$ healthy males aged $36.8 \pm 14.0$ years, $23$ healthy females aged $43.8 \pm 13.6$ years, and $3$ healthy subjects of unknown gender and age.  The AHI and BMI of the male group are $2.7 \pm 1.4$ and $23.1 \pm 2.9$, and the AHI and BMI of the female group are $2.6 \pm 1.4$ and $23.6 \pm 4.6$. The AHI and BMI of the remaining subjects are not known.  CGMH-validation consists of recordings from $10$ healthy males aged $42.3 \pm 12.6$ years, $16$ healthy females aged $47.1 \pm 17.9$ years, and one healthy subject of unknown gender and age.  The AHI and BMI of the male group are $3.0 \pm 1.0$ and $24.4 \pm 3.4$, and the AHI and BMI of the female group are $2.4 \pm 1.6$ and $24.2 \pm 5.0$. The AHI and BMI of the remaining subject are not known. We dismiss $203$ noisy epochs from the training set and $199$ from the testing set because they contain less than five detected R peaks. After dismissal, there are a total of $41,472$ points in the training set and $20,102$ points in the validation set. The percent of wake labels is $18.8\%$ in the training set and $17.1 \%$ in the validation set.
All processing, training, and validation calculations are performed on an Intel\textsuperscript{\textregistered} Core\texttrademark{} $\mathrm{i}7$-$4790\mathrm{K}$ Processor at $3.60$ $\mathrm{GHz}$ with $24$ $\mathrm{GB}$ of RAM and a Microsoft\textsuperscript{\textregistered} Windows\textsuperscript{\textregistered} 10 Home operating system (version 1709). 
In all tables, the positive class corresponds to the ``wake'' label. Because of the imbalance between the positive and negative classes, we should take care when assessing the reported performance measures.

\subsection{Model Validation}

\begin{figure}
\centering
\includegraphics[scale=0.32]{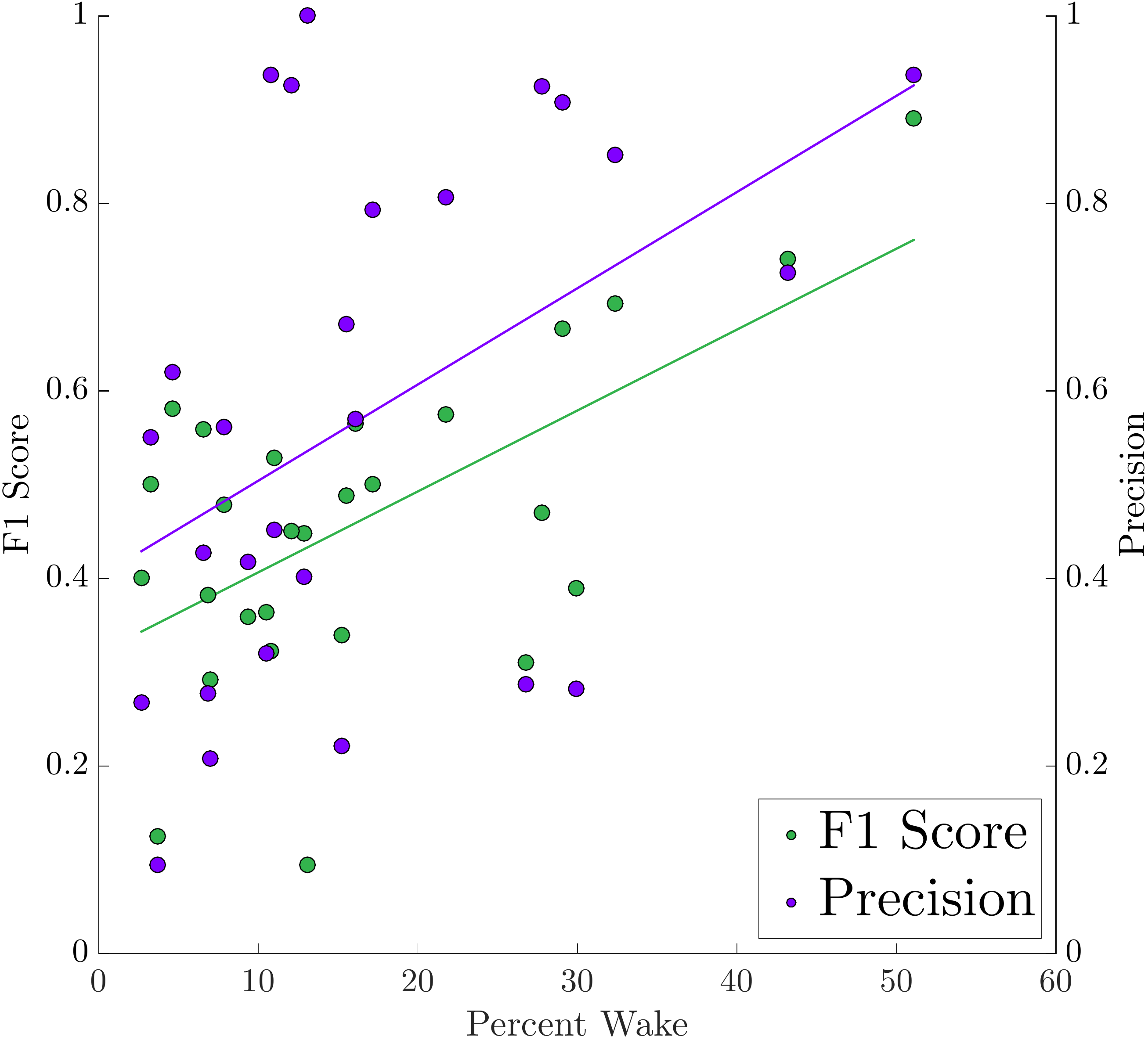}
\caption{\label{Fig:F1PPV} We apply the trained CNN model to each subject in CGMH-validation individually.  We plot the associated $\mathrm{F1}$ and precision values against the percentage of ``wake'' labels in the given recording. We see that the model's performance statistics increase when the percent of ``wake'' labels increases.}
\end{figure}

In Table~\ref{Table:1Q}, we show the performance measures associated with fitting the model to CGMH-training and applying it to the three validation databases. We obtain a good $\mathrm{AUC}$ value of $0.83$ and a moderate Cohen's kappa coefficient of $0.41$ on CGMH-validation. The $\mathrm{F1}$ score is $0.51$, which is not high due to the low precision. To examine the effect of imbalanced classes on the $\mathrm{F1}$ and precision scores, we apply the trained CNN model to each subject in CGMH-validation individually, and we compare the associated $\mathrm{F1}$ and precision values to the percentage of ``wake'' labels in the given recording (see Figure~\ref{Fig:F1PPV}).
If $w$ denotes the percentage of ''wake'' labels in a given recording, a linear regression analysis shows that $\mathrm{F1}=0.32+0.0086w$, where the slope is statistically significant under $t$-test when the significance level is set to $p=0.05$, and $\mathrm{precision} =40.16+1.03w$, where the slope is also statistically significant.

We show that our model performs well when applied to external recordings that are associated with healthy subjects of a different race. The DREAMS Subjects Database has recordings from $16$ healthy females aged $35.8 \pm 15.5$ years and $4$ healthy males as aged $20$, $23$, $27$, and $27$ years.   We dismiss $25$ noisy epochs from the database before evaluation because they are labeled as such or because they contain less than $5$ detected R peaks.  After dismissal, there are $20,032$ epochs remaining, and the percent of wake labels is $16.7 \%$.  The $\mathrm{AUC}$ value is strong at $0.81$, compared to the value $0.83$ on the internal CGMH-validation database.  The sensitivities and specificities are also similar.  The slight difference in performance could be attributed to the fact that the sleep technologists supervising the DREAMS Subjects Database selected recordings which had few artifacts, whereas such artifacts and their physiological correlates were relevant in the CGMH-training data for classification.

\begin{figure}
\centering
\includegraphics[scale=0.32]{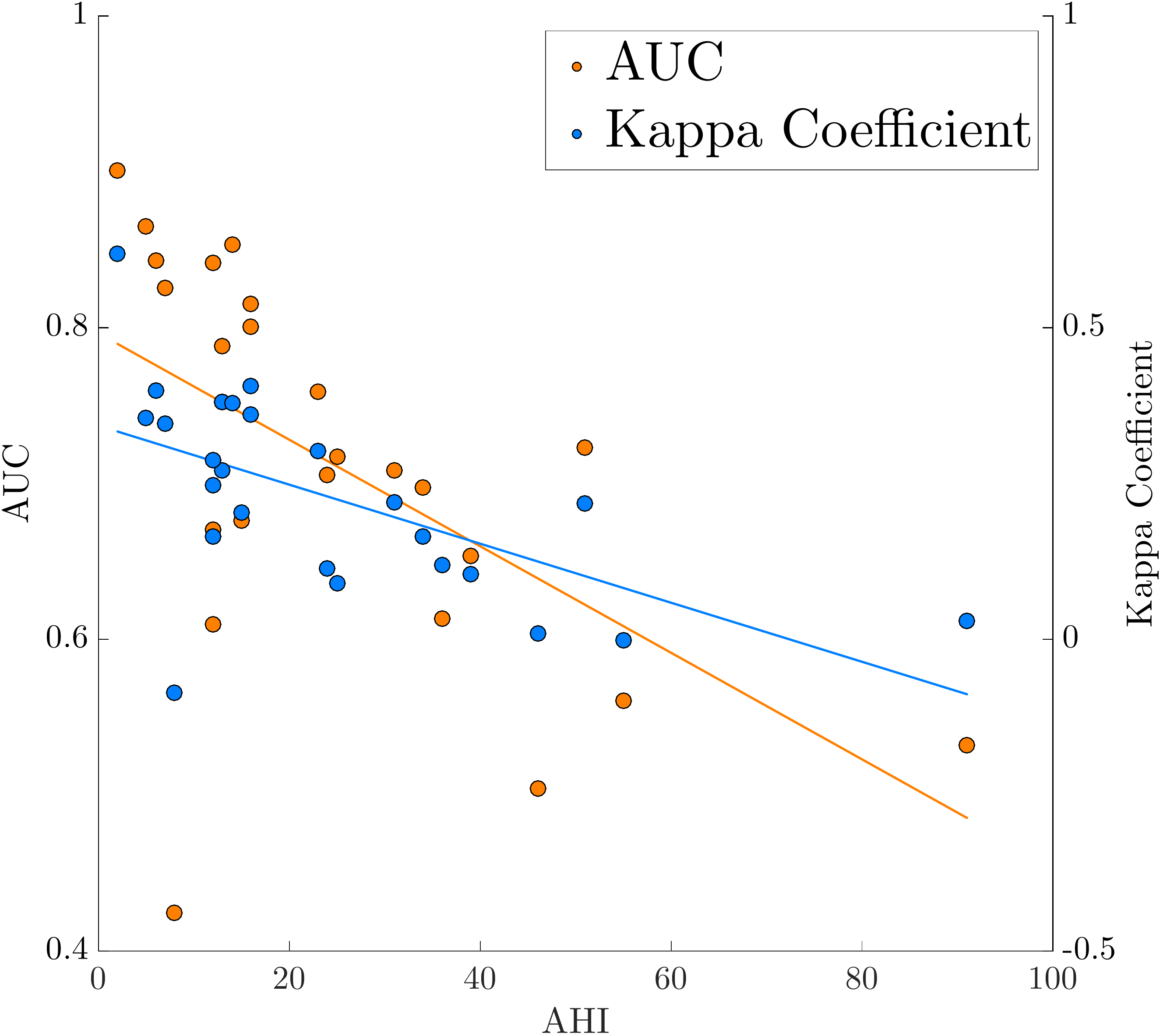}
\caption{\label{Fig:UCDSubjects} We apply the trained CNN model to each subject in the UCDSADB individually.  We plot the associated $\mathrm{AUC}$ and Cohen's kappa values against each subject's apnea-hypopnea index. We see that the model performs well on healthy subjects.}
\end{figure}

We show that our model struggles to achieve convincing results when applied to a database of subjects whose sleep physiology is different. The UCDSADB has recordings from $21$ males aged $48.3 \pm 8.4$ years and $4$ females aged $41$, $62$, $63$, and $68$ years. 
The AHI and BMI of the male group are $25.4 \pm 21.3$ and $31.5 \pm 4.3$, and the AHI and BMI of the female group are $18.3 \pm 14.6$ and $32.2 \pm 2.6$.  
We dismiss $91$ noisy epochs from the database before evaluation because they are labeled as such or because they contain less than $5$ detected R peaks.  After dismissal, there are $19,974$ epochs remaining, and the percent of wake labels is $21.2 \%$.  The performance measures for the UCDSADB, while acceptable, are significantly lower than those observed in the previous model validation steps. To explore the reason for this drop in performance, we evaluate the CNN model on each subject from the UCDSADB individually.  We calculate the $\mathrm{AUC}$ and Cohen's kappa coefficient for each subject and plot these numbers in Figure~\ref{Fig:UCDSubjects} against his or her apnea-hypopnea index (AHI). A subject with a high apnea-hypopnea index experiences frequent interruptions in sleep that arise from breathing difficulties. Since we have trained our model on a database of subjects who do not experience such interruptions, we expect to see a correlation between AHI and model performance measures. We choose $\mathrm{AUC}$ and Cohen's kappa coefficient because they are relatively stable under changes in class size. A linear regression analysis shows that $\mathrm{AUC}=0.8-0.0034\mathrm{AHI}$, where the slope is statistically significant under $t$-test when the significance level is set to $p=0.05$, and $\mathrm{Kappa}=0.34-0.0047\mathrm{AHI}$, where the slope is also statistically significant.

\begin{table}
\caption{Performance statistics for the CNN model trained on CGMH-training}
\label{Table:1Q}
\begin{small}
\begin{center}
\begin{tabular}{c|lcccc}
\hline\\ [-0.7em]
& \multirowcell{2}{CGMH\\-training} & \multirowcell{2}{CGMH\\-validation} & \multirowcell{2}{DREAMS\\Subjects}&\multirowcell{2}{UCDSADB}\\ \\
\hline \\ 
TP   & $4,464$ & $1,800$ & $1,777$ & $1,838$ \\
FP  & $2,143$ & $1,763$ & $2,151$ & $2,853$ \\
TN   & $31,550$ & $14,906$ & $14,532$ & $12,883$\\
FN  & $3,315$ & $1,633$ & $1,572$ & $2,400$\\[0.6em]
SE (\%) & $57.4$ & $52.4$ & 53.1 &  $43.4$\\
SP (\%) & $93.6$ & $89.4$ & 87.1 &  $81.9$\\
ACC (\%) & $86.8$ & $83.1$ & 81.4 &  $73.7$ \\[0.6em]
PR (\%)  & $67.6$ & $50.5$ & 45.2 &  $39.2$\\
F1   & $0.62$ & $0.51$ & 0.49 &  $0.41$\\
AUC   & $0.90$ & $0.83$ & 0.81 &  $0.72$\\
Kappa      & $0.54$ & $0.41$ & 0.38 &  $0.24$ \\
[0.2em]
\hline
\end{tabular}
\end{center}
\end{small}
\vspace{0.1in}
\begin{footnotesize}
TP: true positive; FP: false positive; TN: true negative; FN: false negative; ACC: accuracy; AUC: area under the ROC curve; PR: precision; SE: sensitivity; SP: specificity.
\end{footnotesize}
\end{table}

\subsection{Transfer Proficiency Among Monitoring Devices}

We extract IHR-PPG series from the PPG recordings in CGMH-training and call this database CGMH-training-PPG. Similarly, we extract IHR series from the PPG recordings in CGMH-validation and call this database CGMH-validation-PPG. We then train two models:  one on CGMH-training, and one on CGMH-training-PPG. In Table~\ref{Table:PPG}, we show the performance measures associated with applying both models to CGMH-validation and CGMH-validation-PPG. We see that the overall performance slightly drops when we validate the model on different modality. When the model is both trained and validated on IHR-PPG series, the performance is equivalent, or slightly better, to that of the model trained and validated on the ECG-derived IHR.

\begin{table}
\caption{Transfer proficiency of the CNN model among different monitoring devices}
\label{Table:PPG}
\begin{center}
\begin{small}
\begin{tabular}{c|c cc}
\hline\\ [-0.8em]
\multirowcell{2}{Training\\Database} & \multirowcell{2}{CGMH\\-training} & \multicolumn{2}{c}{\multirow{2}{*}{CGMH-training-PPG}} \\ \\
\hline \\ [-0.8em]
\multirowcell{3}{Validation\\Database} & \multirowcell{3}{CGMH\\-validation\\-PPG}
& \multirowcell{3}{CGMH\\-validation} & \multirowcell{3}{CGMH\\-validation\\-PPG} \\ \\ \\ 
\hline\\ [-0.8em] 
TP   & $2,085$ & $1,217$ & $1,760$\\
FP   & $3,335$ & $1,042$ & $1,522$ \\
TN   & $13,324$ & $15,627$ & $15,137$\\
FN   & $1,331$ & $2,216$ & $1,656$\\[0.6em]
SE (\%) & $61.0$ & $35.5$ & $51.5$\\
SP (\%)  & $80.0$ & $93.8$ & $90.9$ \\ 
ACC (\%)   & $76.8$ & $83.8$ & $84.2$\\[0.6em]
PR (\%)   & $38.5$& $53.9$ & $53.6$ \\
F1    & $0.47$ & $0.43$ & $0.53$\\
AUC     & $0.79$ & $0.81$ & $0.84$ \\
Kappa    & $0.33$ & $0.34$ & $0.43$\\
[0.2em]
\hline
\end{tabular}
\end{small}
\end{center}
\vspace{0.1in}
\begin{footnotesize}
TP: true positive; FP: false positive; TN: true negative; FN: false negative; ACC: accuracy; AUC: area under the ROC curve; PR: precision; SE: sensitivity; SP: specificity.
\end{footnotesize}
\end{table}

\subsection{Sensitivity Analysis}

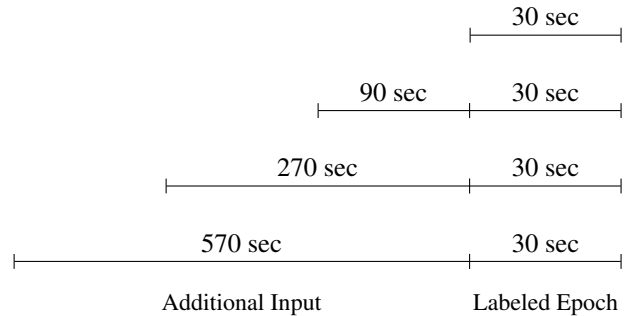
\begin{figure}
\centering
\begin{tikzpicture}
\draw[|-|] (-4,0) -- (4, 0);   
\draw[|-|] (-2,1) -- (4, 1);   
\draw[|-|] (-0,2) -- (4, 2);
\draw[|-|] (2,3) -- (4, 3);     
\draw (2, .1) -- (2, -.1);  
\draw (2, 1.1) -- (2, 1 -.1);  
\draw (2, 2.1) -- (2, 2-.1);  
\node[align=center, above] at (3, 0) {$30 \ \mathrm{sec}$};
\node[align=center, above] at (3, 1) {$30 \ \mathrm{sec}$};
\node[align=center, above] at (3, 2) {$30 \ \mathrm{sec}$};
\node[align=center, above] at (3, 3) {$30 \ \mathrm{sec}$};
\node[align=center, below=0.3] at (3, 0) {\small Labeled Epoch};
\node[align=center, below=0.3] at (-1, 0) {\small Additional Input};
\node[align=center, above] at (1, 2) {$90 \ \mathrm{sec}$};
\node[align=center, above] at (0, 1) {$270 \ \mathrm{sec}$};
\node[align=center, above] at (-1, 0) {$570 \ \mathrm{sec}$};
\end{tikzpicture}
\caption{\label{Fig:EpochLength} We construct input signals of various lengths by additionally considering data which precedes the labeled epoch.}
\end{figure}

We examine the effect of input size on the performance of the CNN model. (See Figure~\ref{Fig:EpochLength} for a depiction of the considered input sizes.) In Table~\ref{Table:1}, we show the performance measures associated with fitting the model to CGMH-training and applying it to the three validation databases. When the model is trained using an input size of $30$ seconds, the validation performance is lower than when the model is trained using any amount of the preceding information. When the input size is $2$ minutes, the performance is closer to the performance obtained using an input size of $5$ minutes. Note, however, that the benefits of including large amounts of preceding information are not unlimited.  We observe no significant improvement in validation $\mathrm{AUC}$ by considering $10$ minutes instead of $5$ minutes, whereas the improvement obtained by considering $5$ minutes instead of $30$ seconds is large. Evidently, the lack of improvement when considering an input size of $10$ minutes is the result of over-fitting to the training database.
Whereas the training process for input signals of length $30$ seconds takes approximately $20$ minutes, the same process takes over $5$ hours for input signals of length $10$ minutes. We declare a trade-off between computational efficiency and classification proficiency.  

\begin{table}
\caption{Sensitivity analysis for the CNN model trained on CGMH-training with various input sizes}
\label{Table:1}
\begin{center}
\begin{small}
\begin{tabular}{c|lccc}
\hline\\ [-0.7em]
 & Input Size & \textsc{$30$ Sec} & \textsc{$2$ Min}  & \textsc{$10$ Min}\\ [0.2em]
\hline \\ [-0.6em]  \multirowcell{11}{CGMH\\-training}
&TP   & $2,860$ & $4,240$ &  $6,387$ \\
&FP  & $1,590$ & $1,416$ & $2,419$ \\
&TN   & $32,127$ & $32,294$  & $31,198$\\
&FN  & $5,399$ & $3,858$ &  $908$\\[0.6em]
&SE (\%)  & $34.6$ & $52.4$ &   $87.6$\\
&SP (\%) & $95.3$ & $95.8$ &   $92.8$\\
&ACC (\%)   & $83.4$ & $87.4$ &   $91.9$ \\[0.6em]
&PR (\%)  & $64.3$ & $75.0$ &   $74.3$\\
&F1    & $0.45$ & $0.61$ &   $0.79$\\
&AUC        & $0.81$ & $0.90$  &  $0.96$\\
&Kappa      & $0.36$ & $0.54$ &   $0.74$ \\[0.2em]
\hline\\ [-0.6em] \multirowcell{11}{CGMH\\-validation}
&TP  & $1,067$ & $1,619$     & $2,012$ \\
&FP  & $1,164$ & $1,705$   & $2,527$ \\
&TN   & $15,520$ & $14,977$     & $14,072$\\
&FN  & $2,594$ & $1,963$  & $1,221$\\[0.6em]
&SE (\%) & $29.2$ & $45.2$ &  $62.2$\\
&SP (\%) & $93.0$ & $89.9$ &  $84.8$\\
&ACC (\%)   & $81.5$ & $81.9$ &  $81.1$ \\[0.6em]
&PR (\%)  & $47.8$ & $48.7$ &  $44.3$\\
&F1    & $0.36$ & $0.47$ &  $0.52$\\
&AUC        & $0.75$ & $0.81$ &  $0.83$\\
&Kappa      & $0.26$ & $0.36$ &  $0.40$ \\
[0.2em]
\hline\\ [-0.6em] \multirowcell{11}{DREAMS\\Subjects}
&TP   & $998$ & $1,458$  & $1,775$ \\
&FP  & $1,197$ & $1,891$  & $2,267$ \\
&TN   & $15,486$ & $14,792$  & $14,404$\\
&FN  & $2,531$ & $2,011$  & $1,386$\\[0.6em]
&SE (\%) & $28.3$ & $42.0$  &  $56.2$\\
&SP (\%) & $92.8$ & $88.7$ &  $86.4$\\
&ACC (\%)   & $81.6$ & $80.6$  &  $81.6$ \\[0.6em]
&PR (\%) & $45.5$ & $43.5$  &  $43.9$\\
&F1   & $0.35$ & $0.43$  &  $0.49$\\
&AUC        & $0.69$ & $0.76$  &  $0.81$\\
&Kappa      & $0.25$ & $0.31$  &  $0.38$ \\
[0.2em]
\hline\\ [-0.6em] \multirow{11}{*}{UCDSADB}
&TP  & $1,344$ & $1,789$    & $2,199$ \\
&FP  & $2,293$ & $2,874$   & $3,963$ \\
&TN   & $13,446$ & $12,865$     & $11,750$\\
&FN  & $3,116$ & $2,596$  & $1,812$\\[0.6em]
&SE (\%) & $30.1$ & $40.8$ &  $54.8$\\
&SP (\%) & $85.4$ & $81.7$ &  $74.8$\\
&ACC (\%)   & $73.2$ & $72.8$ &  $70.7$ \\[0.6em]
&PR (\%)  & $37.0$ & $38.4$ &  $35.7$\\
&F1   & $0.33$ & $0.39$ &  $0.43$\\
&AUC        & $0.61$ & $0.68$ &  $0.69$\\
&Kappa      & $0.17$ & $0.22$ &  $0.25$ \\
[0.2em]
\hline
\end{tabular}
\end{small}
\end{center}
\vspace{0.1in}
\begin{footnotesize}
TP: true positive; FP: false positive; TN: true negative; FN: false negative; ACC: accuracy; AUC: area under the ROC curve; PR: precision; SE: sensitivity; SP: specificity.
\end{footnotesize}
\vskip -0.1in
\end{table}

Our sensitivity analysis proceeds. We fit the CNN model to our external validation databases, one by one, so that we have two trained models. We use an input size of $5$ minutes. In Table~\ref{Table:Mix}, we report the performance measures associated with applying these models to our three validation databases. Note that some of the performance measures are inflated because the model has been applied to its training database.
We notice that the DREAMS Subjects Database does not provide a result which generalizes well to all other databases. Specifically, the associated model obtained an $\mathrm{AUC}$ value of $0.67$ on the UCDSADB. This poor performance could be attributed to the size of the DREAMS Subjects Database or the relative homogeneity of the subjects compared to the UCDSADB.  However, the model trained on the UCDSADB does transfer effectively to the other databases, and we suggest that the reason involves the fact that number of ``wake'' labels is relatively high.  This result is surprising because the number of subjects in this database is less than half the number of subjects in CGMH-training.

\begin{table}
\caption{Performance statistics for the CNN models trained on the DREAMS Subjects Database and the UCDSADB}
\label{Table:Mix}
\begin{center}
\begin{small}
\begin{tabular}{c|lccc}
\hline\\ [-0.7em]
\multirowcell{2}{\diagbox[width=7em]{Training}{Validation}} & \multirowcell{2}{} & \multirowcell{2}{CGMH\\-validation} & \multirowcell{2}{DREAMS\\Subjects} & \multirowcell{2}{UCDSADB}\\ \\ [0.2em]
\hline\\ [-0.6em] \multirowcell{11}{DREAMS\\Subjects}
&TP   & $1,517$ & $2,095$  & $1,166$ \\
&FP  & $1,928$ & $1,396$  & $1,713$ \\
&TN   & $14,741$ & $15,287$ & $14,023$\\
&FN  & $1,916$ & $1,254$  & $3,072$ \\ [0.6em]
&SE & $44.2$ & $62.6$ &  $27.5$\\
&SP & $88.4$ & $91.6$ &  $89.1$\\
&ACC   & $80.9$ & $86.8$ &  $76.0$ \\[0.6em]
&PR  & $44.0$ & $60.0$ &  $40.5$\\
&F1    & $0.44$ & $0.61$ &  $0.33$\\
&AUC        & $0.77$ & $0.89$ &  $0.67$\\
&Kappa      & $0.33$ & $0.53$ &  $0.19$ \\
[0.2em]
\hline\\ [-0.6em] \multirowcell{11}{UCDSADB}
&TP   & $1,579$ & $1,723$ & $2,169$ \\
&FP  & $2,271$ & $3,000$ & $911$ \\
&TN   & $14,398$ & $13,706$ & $14,825$ \\
&FN  & $1,854$ & $1,626$ & $2,069$ \\[0.6em]
&SE & $46.0$ & $51.5$ & $51.2$ \\
&SP & $86.4$ & $82.0$ & $94.2$\\
&ACC   & $79.5$ & $76.9$ & $85.1$  \\[0.6em]
&PR  & $41.0$ & $36.5$ & $70.4$ \\
&F1    & $0.43$ & $0.43$ & $0.59$ \\
&AUC        & $0.75$ & $0.75$ & $0.87$ \\
&Kappa      & $0.31$ & $0.29$ & $0.50$ \\
[0.2em]
\hline
\end{tabular}
\end{small}
\end{center}
\vspace{0.1in}
\begin{footnotesize}
TP: true positive; FP: false positive; TN: true negative; FN: false negative; ACC: accuracy; AUC: area under the ROC curve; PR: precision; SE: sensitivity; SP: specificity.
\end{footnotesize}
\vskip -0.1in
\end{table}

\section{Further exploration of CNN via Data and Feature Visualization}\label{Sect:Exploration}

\begin{figure*}
\centering
\includegraphics[scale=0.2]{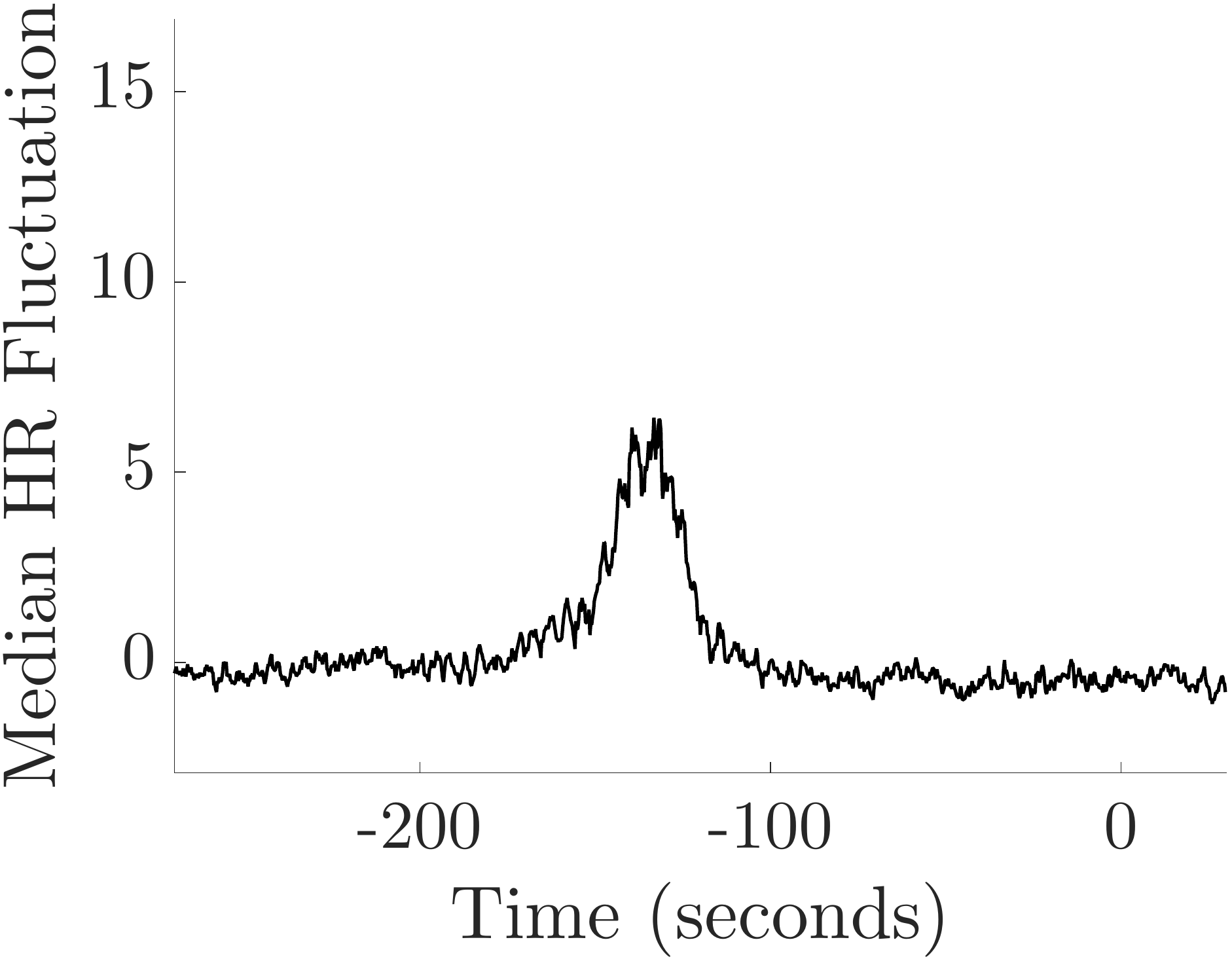}\includegraphics[scale=0.2]{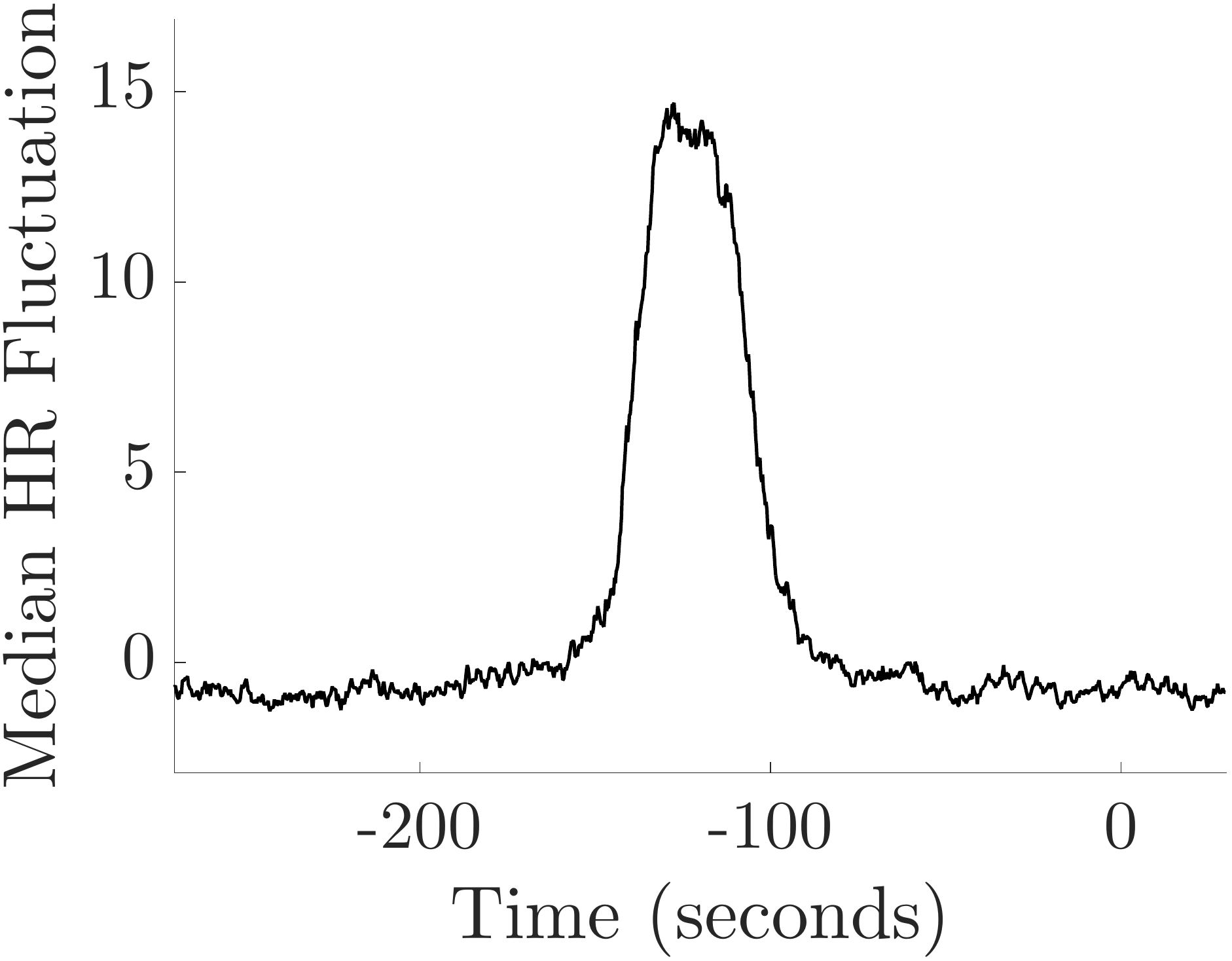}\includegraphics[scale=0.2]{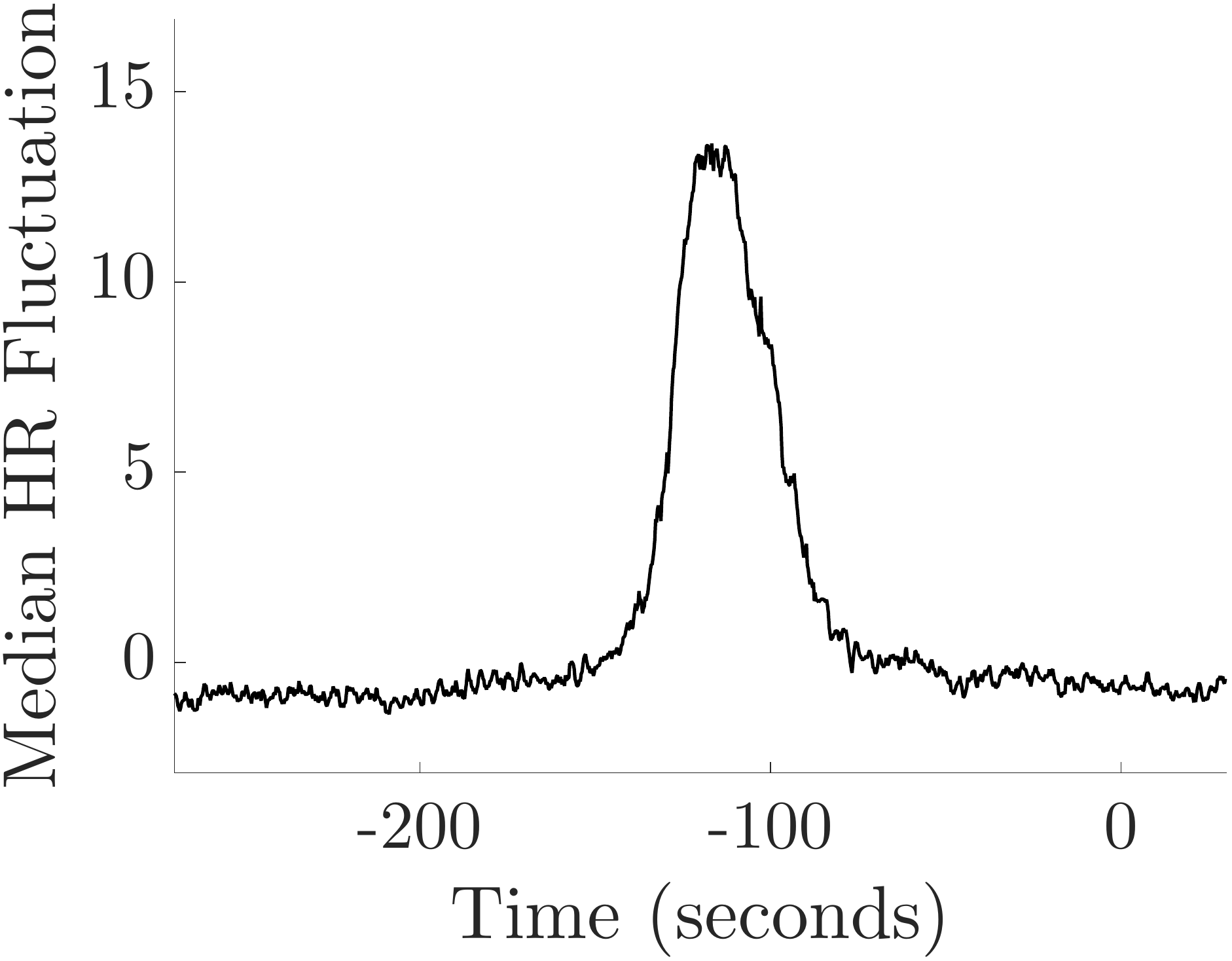}\includegraphics[scale=0.2]{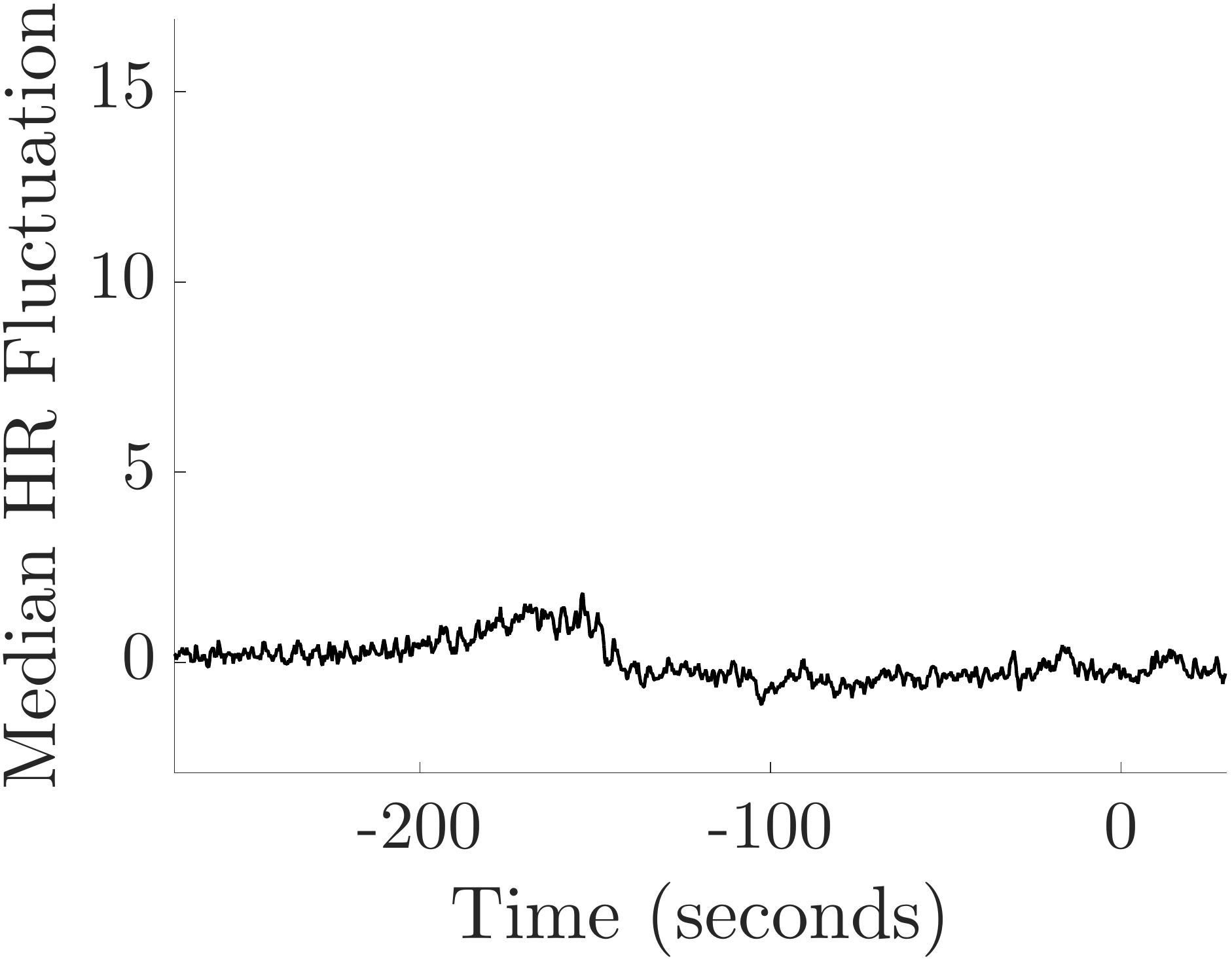}\\
\includegraphics[scale=0.2]{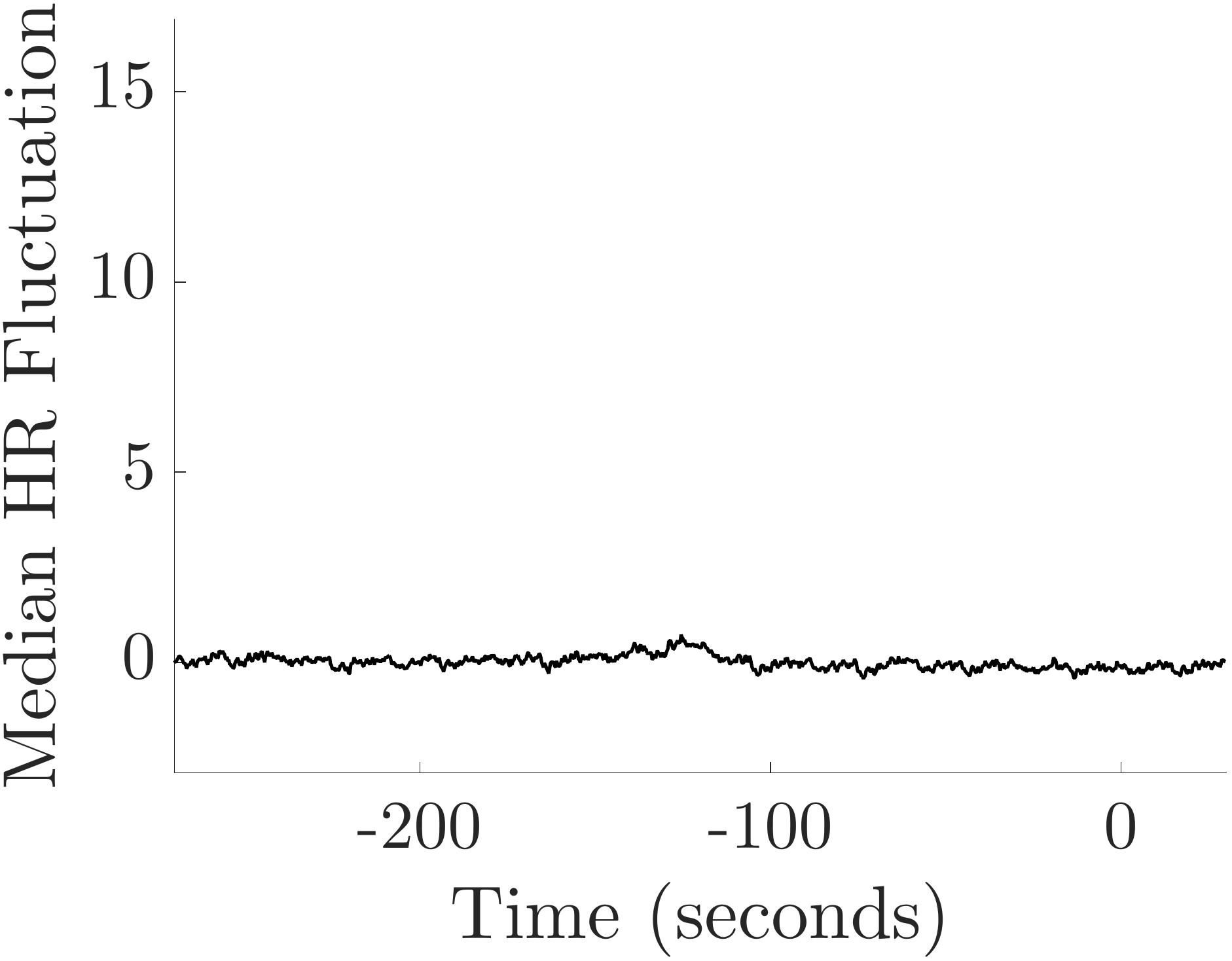}\includegraphics[scale=0.2]{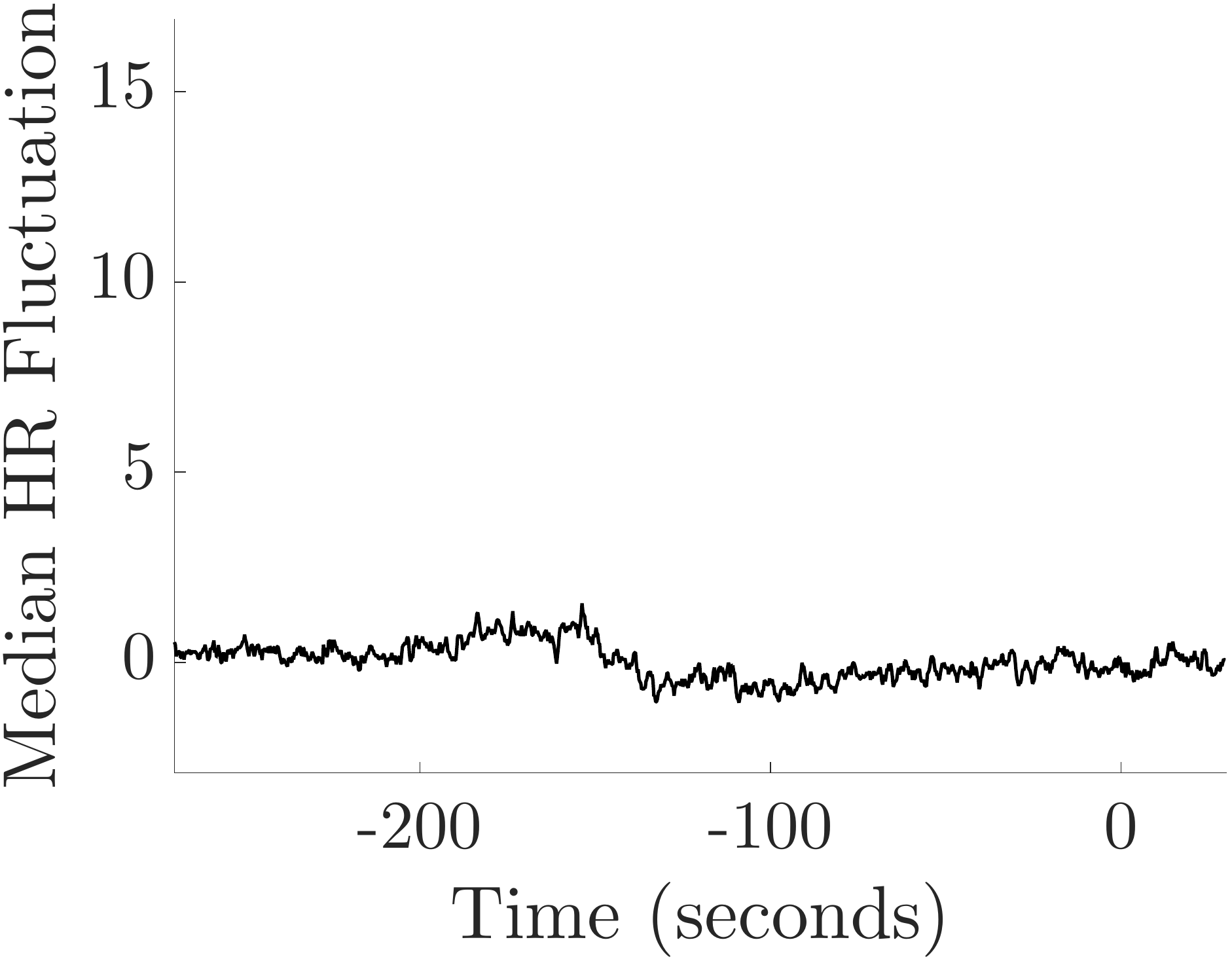}\includegraphics[scale=0.2]{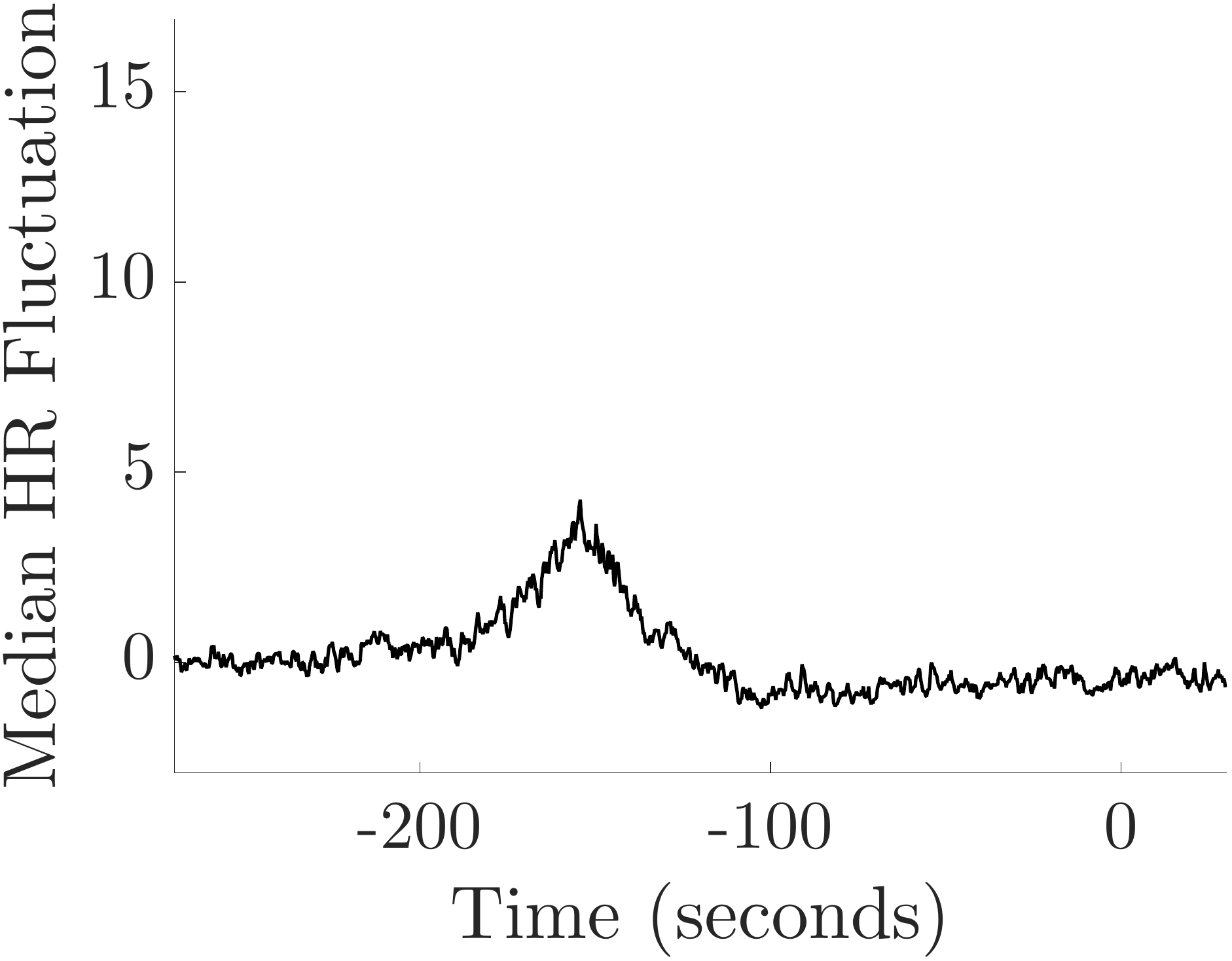}\includegraphics[scale=0.2]{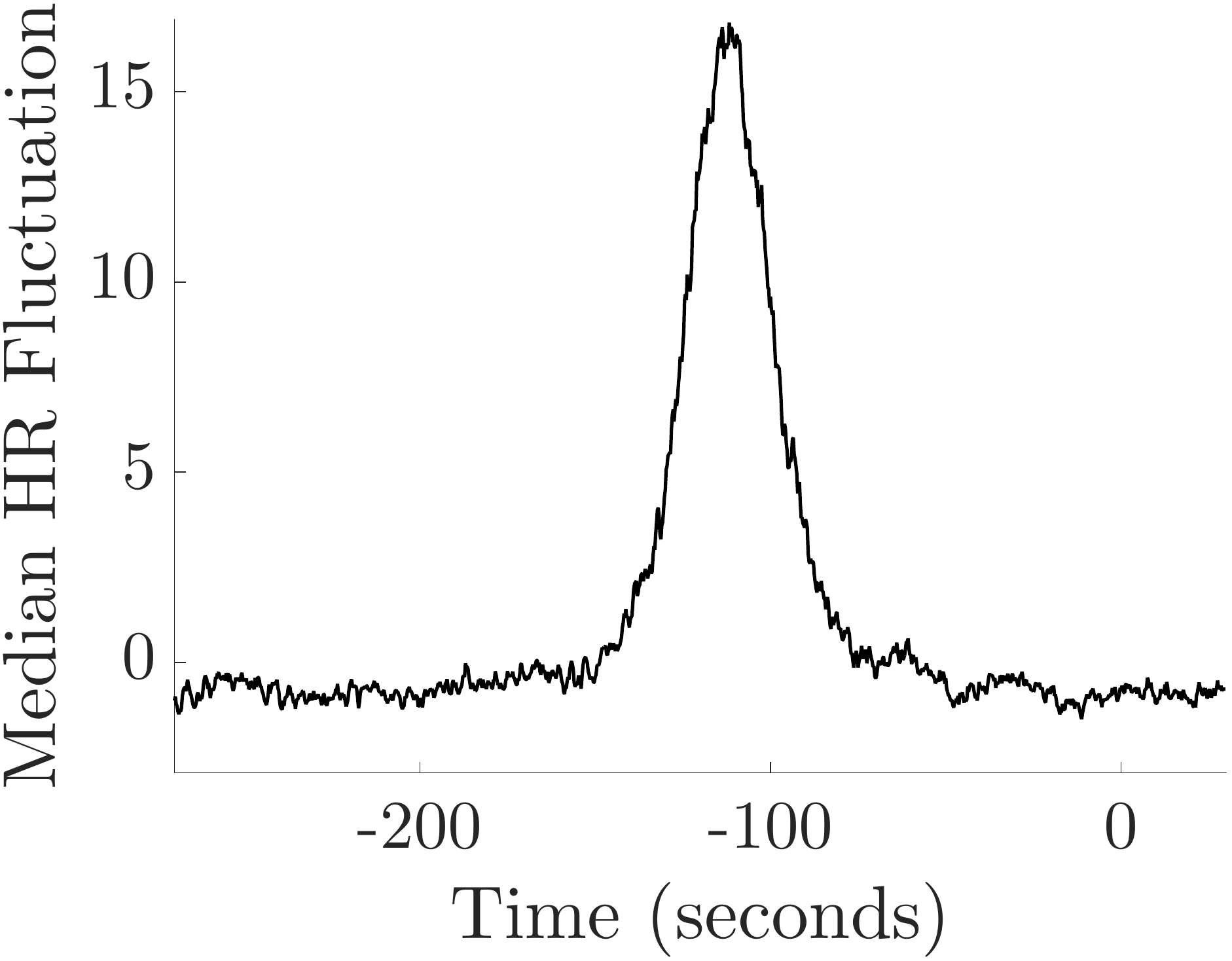}\\
\includegraphics[scale=0.2]{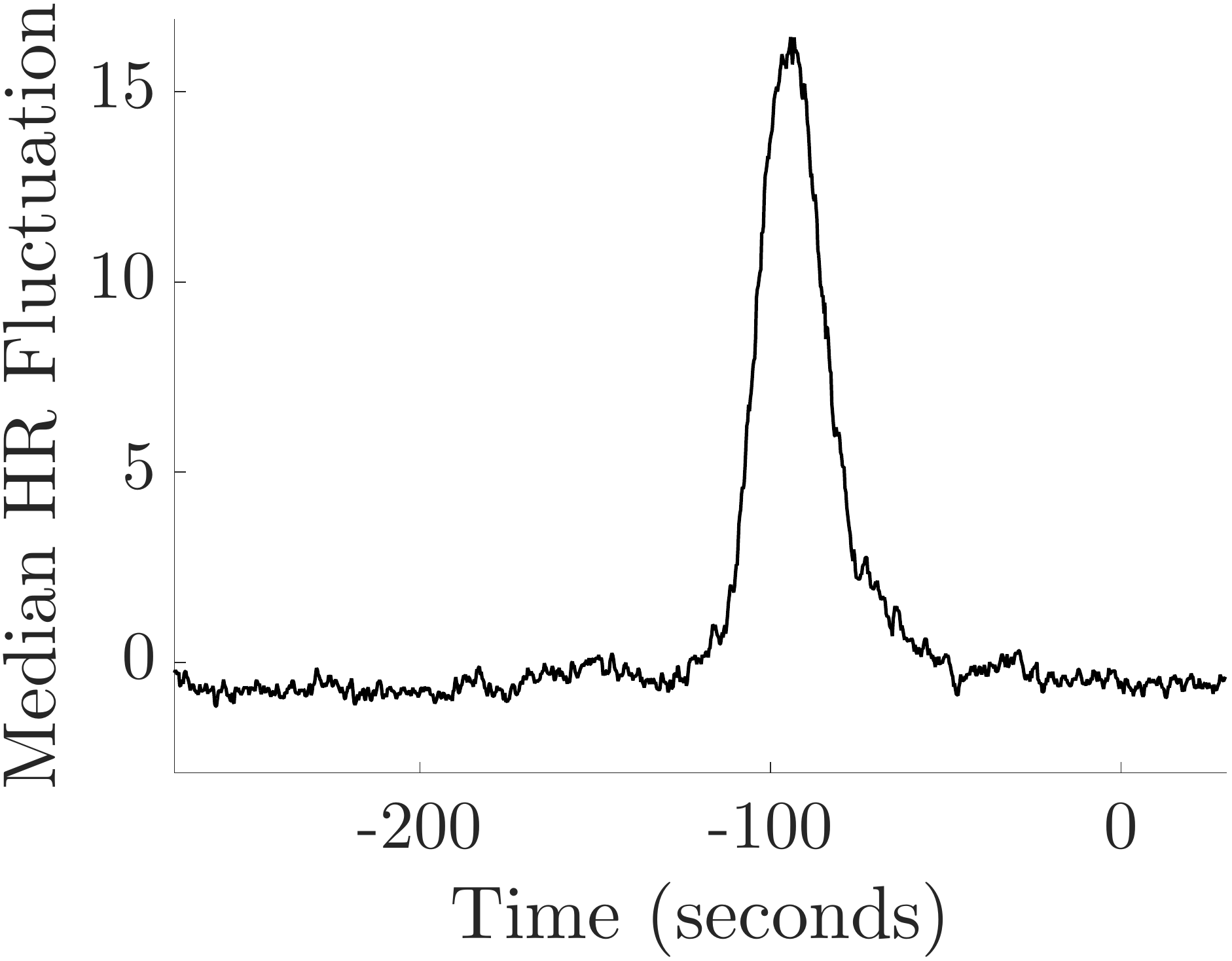}\includegraphics[scale=0.2]{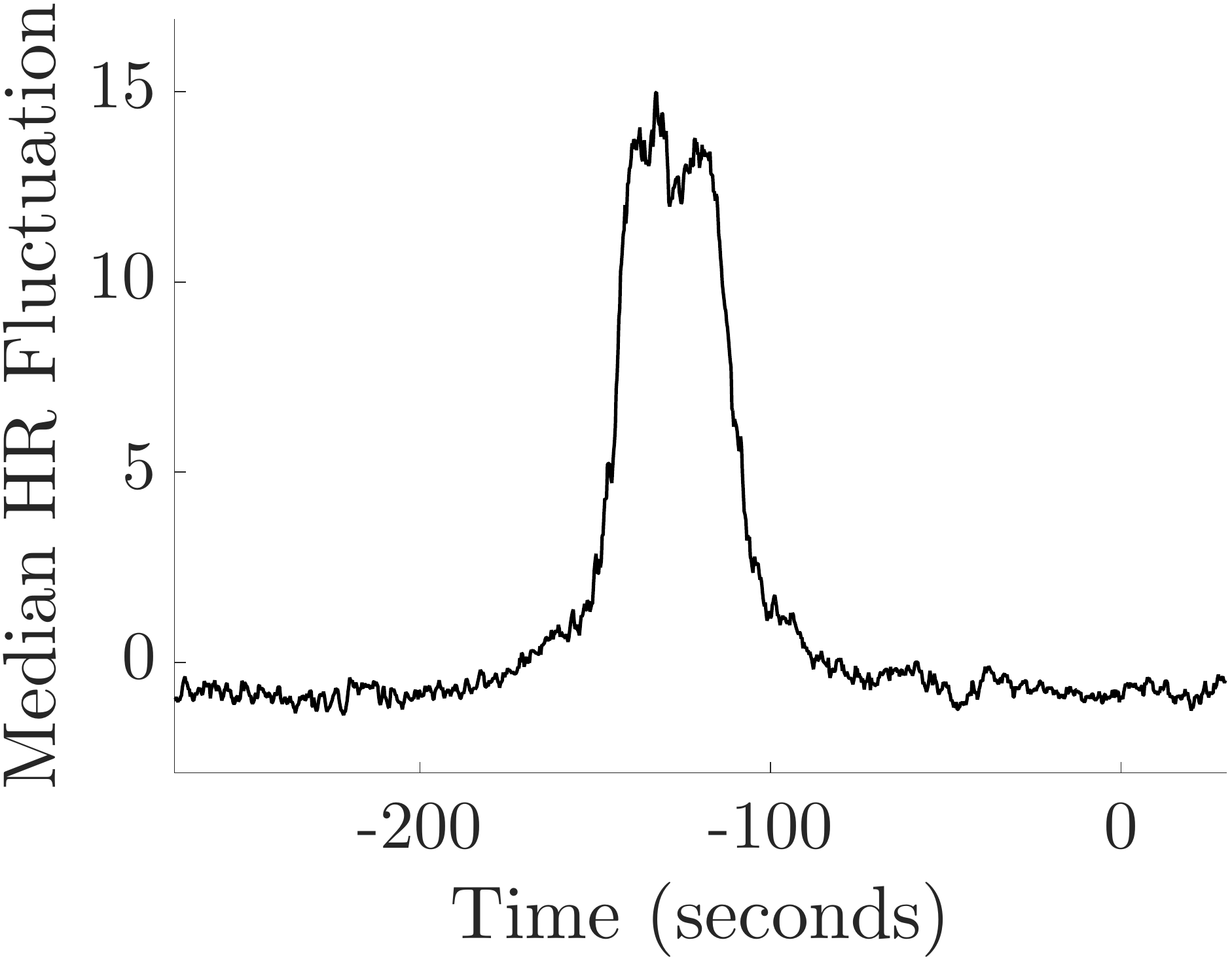}
\caption{\label{Fig:Features}We illustrate the features captured by the trained CNN model. The plots in the first row are $\psi_1(18)$-$\psi_4(18)$, the plots in the second row are $\psi_5(18)$-$\psi_8(18)$, and the plots in the third row are $\psi_9(18)$-$\psi_{10}(18)$.}
\end{figure*}

Although there have been several works trying to theoretically understand CNNs, like \cite{Mallat2012,LinRolnick2017,Wiatowski2017}, our knowledge about the CNN framework is still limited. We should take care to observe that the behaviour of our model is not unreasonable. At the very least, viewing the activity of the model provides insight into how the network captures HRV quantities. We introduce some notation to ease the exposition. Write $\{x_i\}_{i=1}^{20102} \subset \mathbb{R}^{1200}$ for the $5$-minute input signals from CGMH-validation.  For each $i$, let $y_i = f(x_i) \in \mathbb{R}^{38 \times 10}$ denote the output of the last convolution block in the network, where $f \colon \mathbb{R}^{1200} \rightarrow \mathbb{R}^{38 \times 10}$ is used to denote applying the convolutional section of the network in a feed-forward fashion. The size of each $y_i$ is determined by the network architecture:  there are $10$ filters in the last convolutional layer, and we think of the $38 \times 10$ matrix $y_i$ as $10$ filtered and down-sampled versions of the original $x_i$. The intuition is that if $y_i^j \in \mathbb{R}^{38}$ denotes the $j^\text{th}$ column of $y_i$, i.e. the $j^\text{th}$ filtered and down-sampled version of $x_i$, then the probability that $x_i$ contains some feature $\psi_j$ at a time corresponding to the $t^\text{th}$ sample $(1 \leq t \leq 38)$ is $y_i^j(t)$.  Note that $f$ is a non-negative matrix-valued function because of the ReLU activation function, which maps all negative numbers to zero.  We discover the morphology of the features $\psi_j$ as follows. Fix one of the $38$ samples $t$,
%(e.g. $-120 \ \mathrm{sec}$)
and fix $j \in \{1, ..., 10\}$, one of the $10$ non-linearly filtered signals outputted by the final convolution block. Consider the subset of $400$ indices defined sequentially as
\begin{gather}
i_k := \operatorname*{arg\,max} \{y_i^j(t) : i \notin \{i_1, ..., i_{k-1}\} \} \quad 1 \leq k \leq 400,
\end{gather}
and set
\begin{gather}
\psi_j(t) := \operatorname*{median} \{x_{i_k} : 1 \leq k \leq 400 \} \in \mathbb{R}^{1200}.
\end{gather}
In Figure~\ref{Fig:Features}, we plot $\psi_j(t)$ for $1 \leq j \leq 10$ and $t = 18$. These plots represent the common information held among input signals with similar activations. Note that choosing a different sample $t$ centers the main activity of the plot at a different point in time.
The captured features appear to be related to low-frequency variabilities inside the IHR series, but this result could be merely a consequence of the fact that the averaging process cancels out the detailed information.
 
\begin{figure*}
\centering
\subfigure[Features indicative of the wake stage]{\includegraphics[scale=0.32]{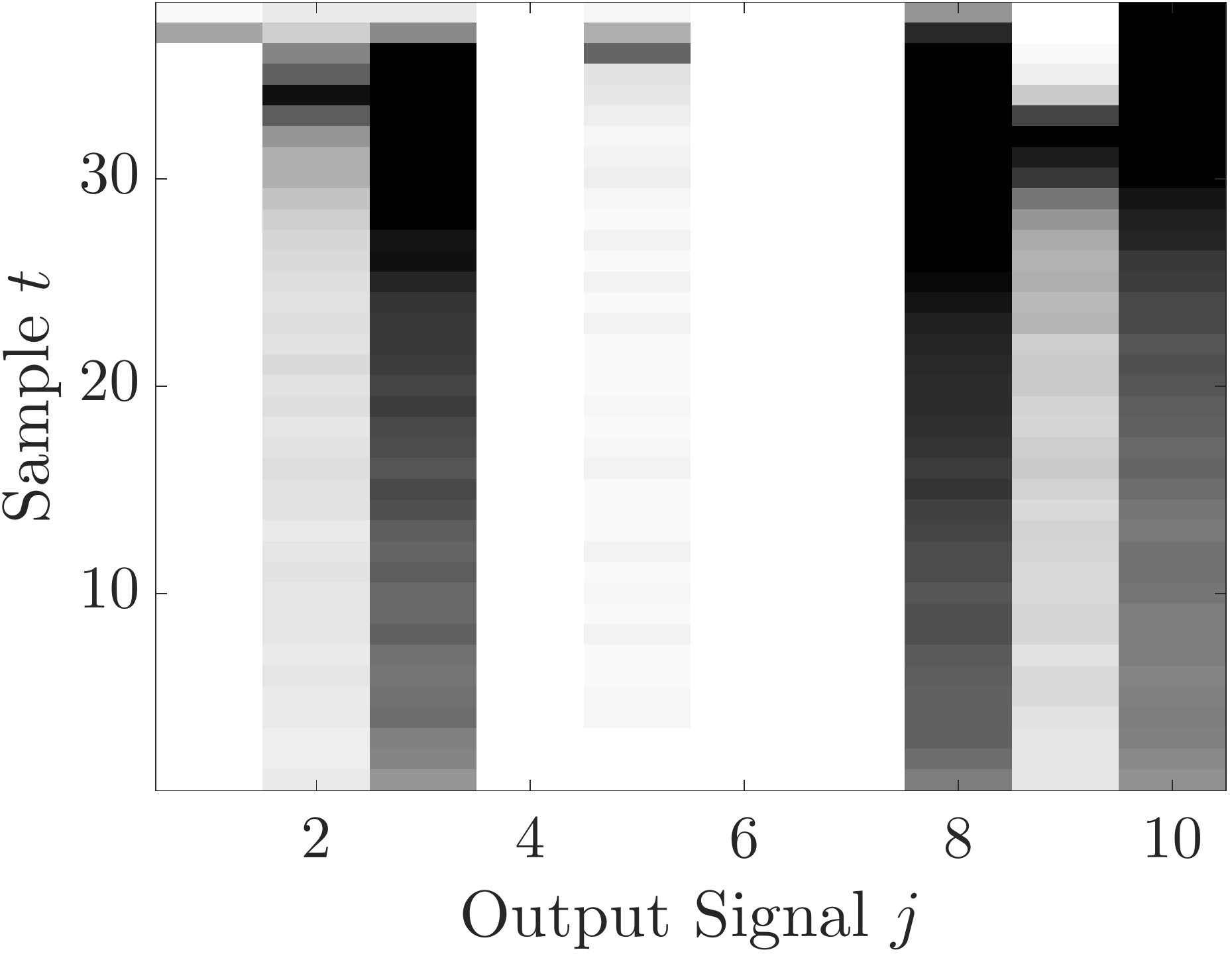}}
\hspace{0.5in} \subfigure[Features indicative of the sleep stage]{\includegraphics[scale=0.32]{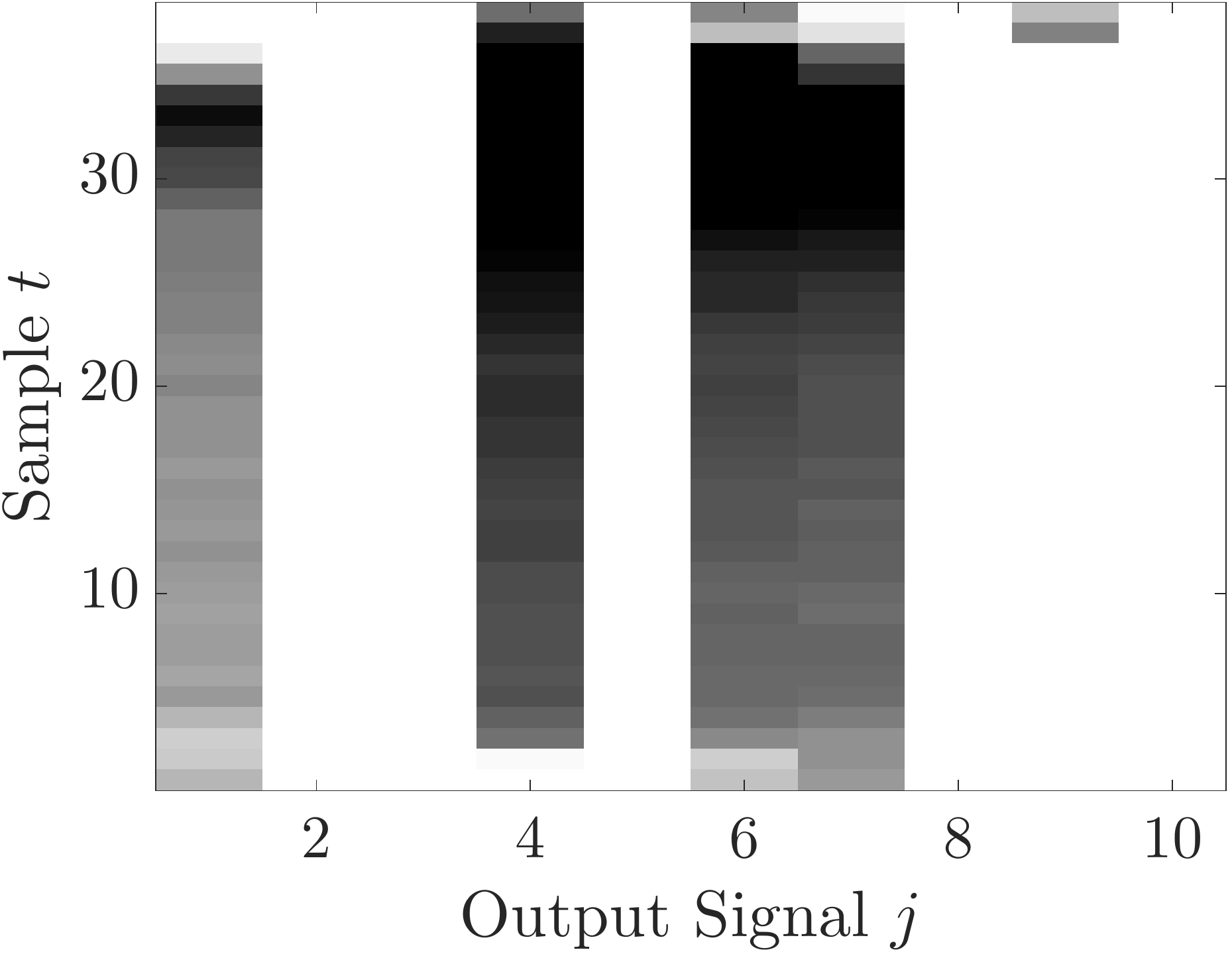}}
\caption{\label{Fig:pValue} At each time $t$, we assess the predictive potential of the $10$ features $\psi_j$ learned by the CNN model. The black colour in the $(j, t)$-entry of (a) (resp. (b)) indicates that the feature $\psi_j$ is useful for identifying the wake (resp. sleep) stage when it occurs at sample $t$.}
\end{figure*}

 According to our analysis, the features in plots $2$, $3$, $5$, $8$, $9$, and $10$ are indicative of the wake stage, whereas the other plots are indicative of the sleep stage.  The procedure is as follows. For each sample $t \in \{1, ..., 38\}$ and each output signal $j \in \{1, ..., 10\}$, we perform two statistical tests using the labeled data from CGMH-validation.  For each $i$, let $x_i(l)$ denote the label ($1$ or $0$ for ``wake'' or ``sleep'', respectively) associated to the input signal $x_i$. In Figure~\ref{Fig:pValue} (a), the $(j, t)$-entry is ${-\log} p$, where $p$ is the probability that the set
\begin{gather}
 \{y_i^j(t) : x_i(l) = 1\} 
 \end{gather}
 is sampled from a distribution with median less or equal to the median of the distribution from which the set
 \begin{gather}
 \{y_i^j(t) : x_i(l) = 0\}
 \end{gather}
 is sampled. The colour black is associated with large values of $-\log p$ and the proposition that if $x_i$ possesses feature $\psi_j$ at a time corresponding to sample $t$, then $x_i(l)$ is likely equal to $1$. In Figure~\ref{Fig:pValue} (b), the $(j, t)$-entry is ${-\log} p$, where $p$ is the probability that the set
\begin{gather}
 \{y_i^j(t) : x_i(l) = 0\} 
 \end{gather}
 is sampled from a distribution with median less or equal to the median of the distribution from which the set
 \begin{gather}
 \{y_i^j(t) : x_i(l) = 1\}
 \end{gather}
 is sampled. The colour black is associated with large values of $-\log p$ and the proposition that if $x_i$ possesses feature $\psi_j$ at a time corresponding to sample $t$, then $x_i(l)$ is likely equal to $0$. Remember that the label associated to $x_i$ is based on the last $30$ seconds of $x_i$.  What we see is expected: detecting a feature closer to the end of the signal ($t \rightarrow 38$) is more useful for determining the true label of the input signal. 
 
\begin{figure*}
\centering
\vspace{0.2in}
\includegraphics[scale=0.2]{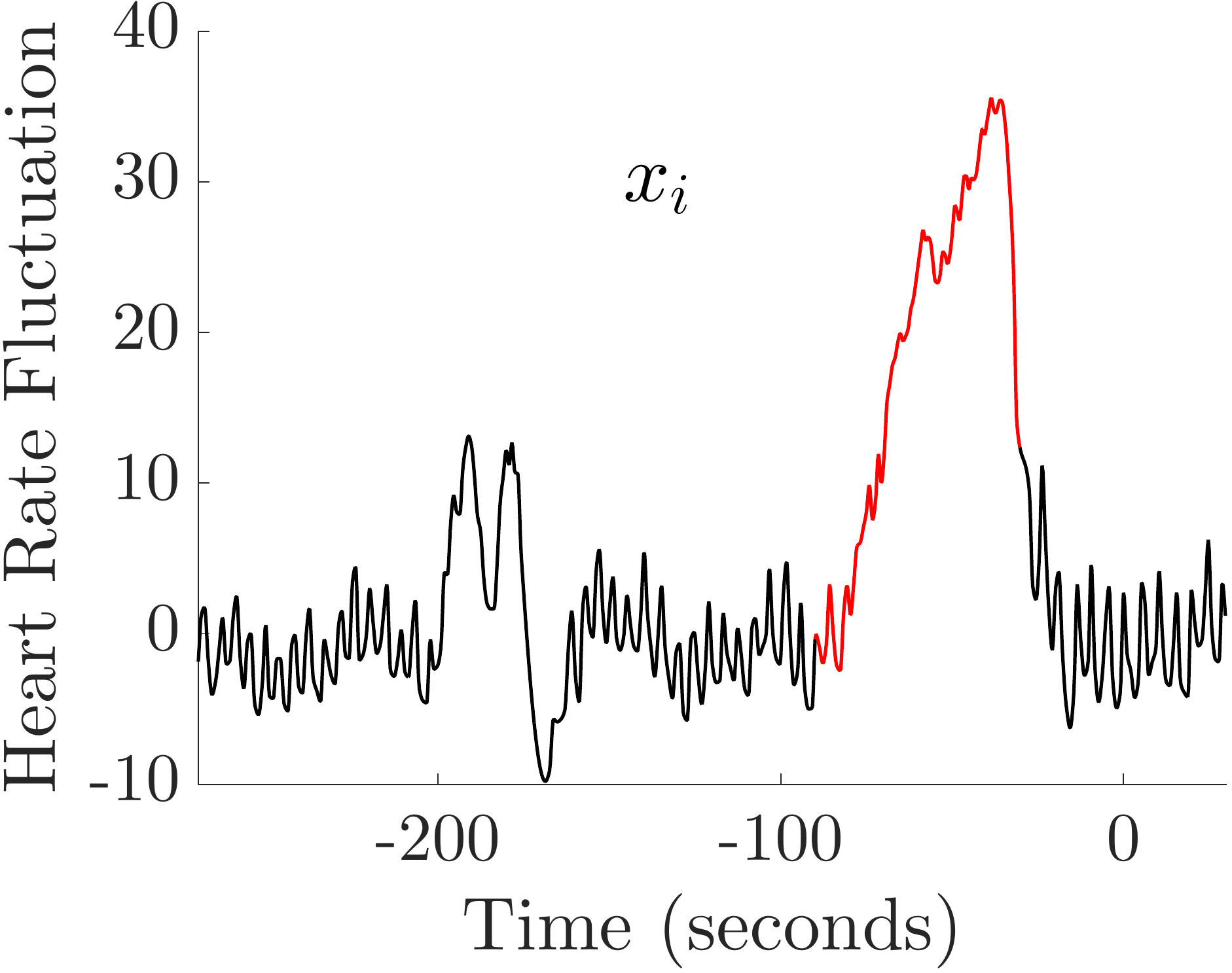}\includegraphics[scale=0.2]{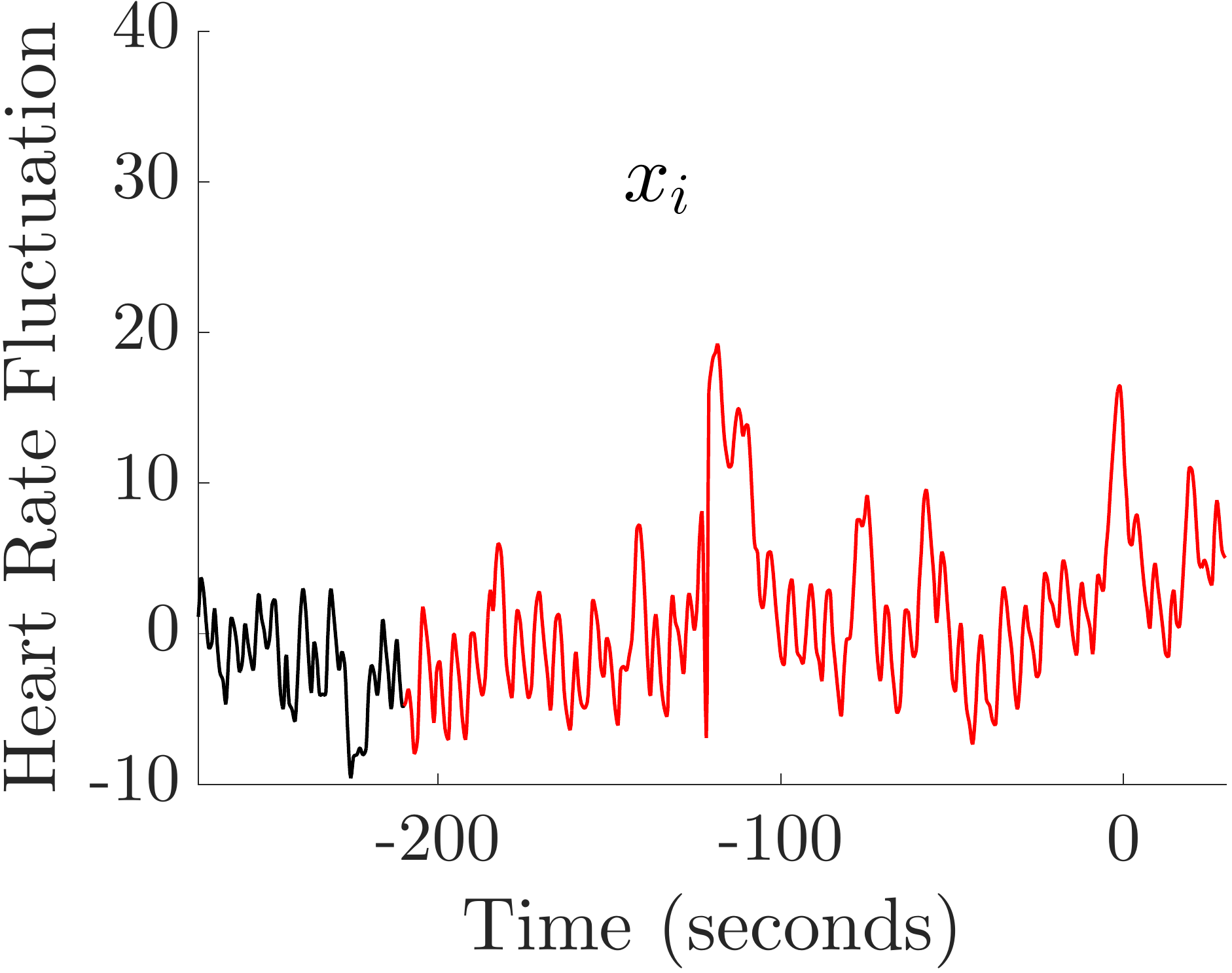}\includegraphics[scale=0.2]{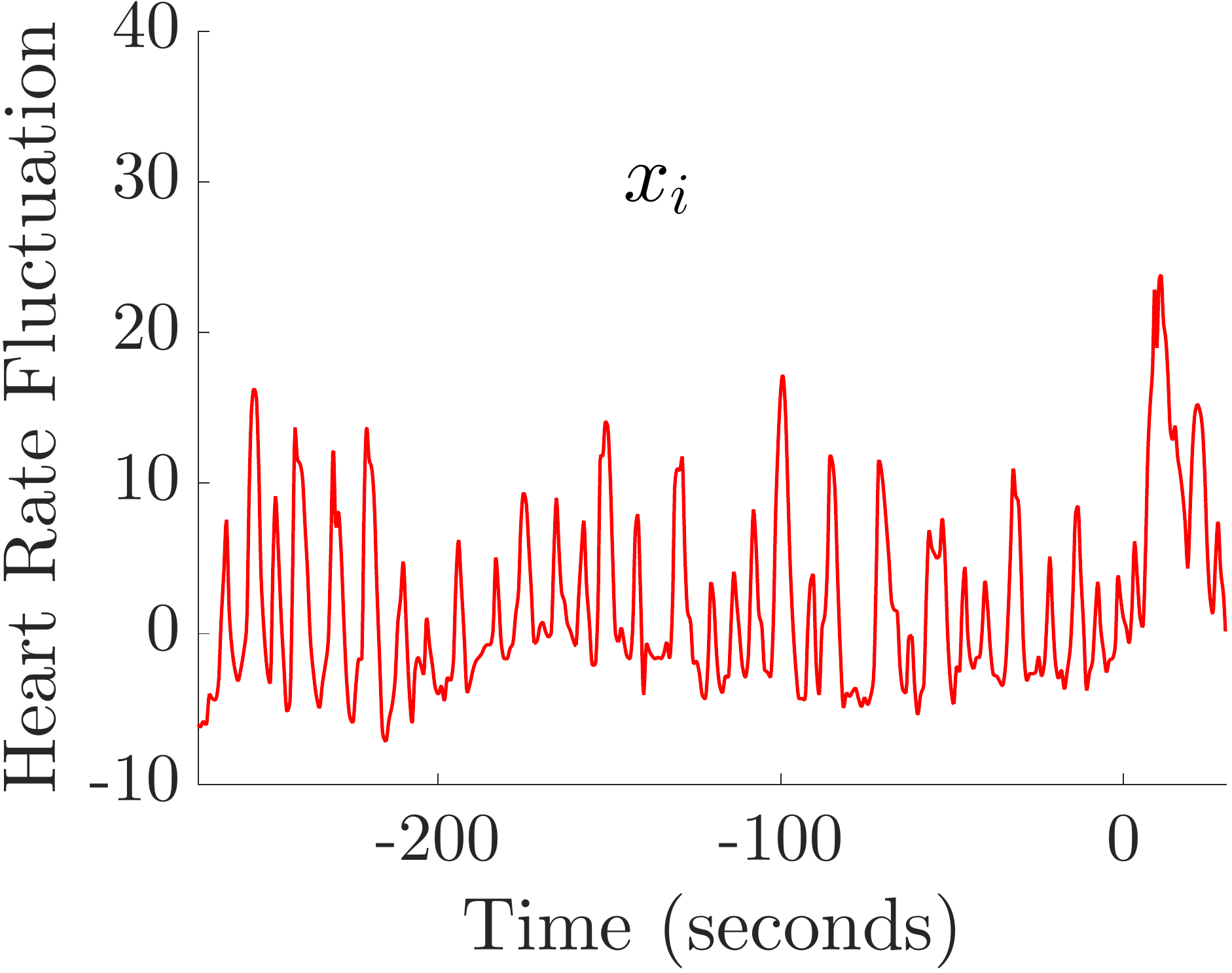}\includegraphics[scale=0.2]{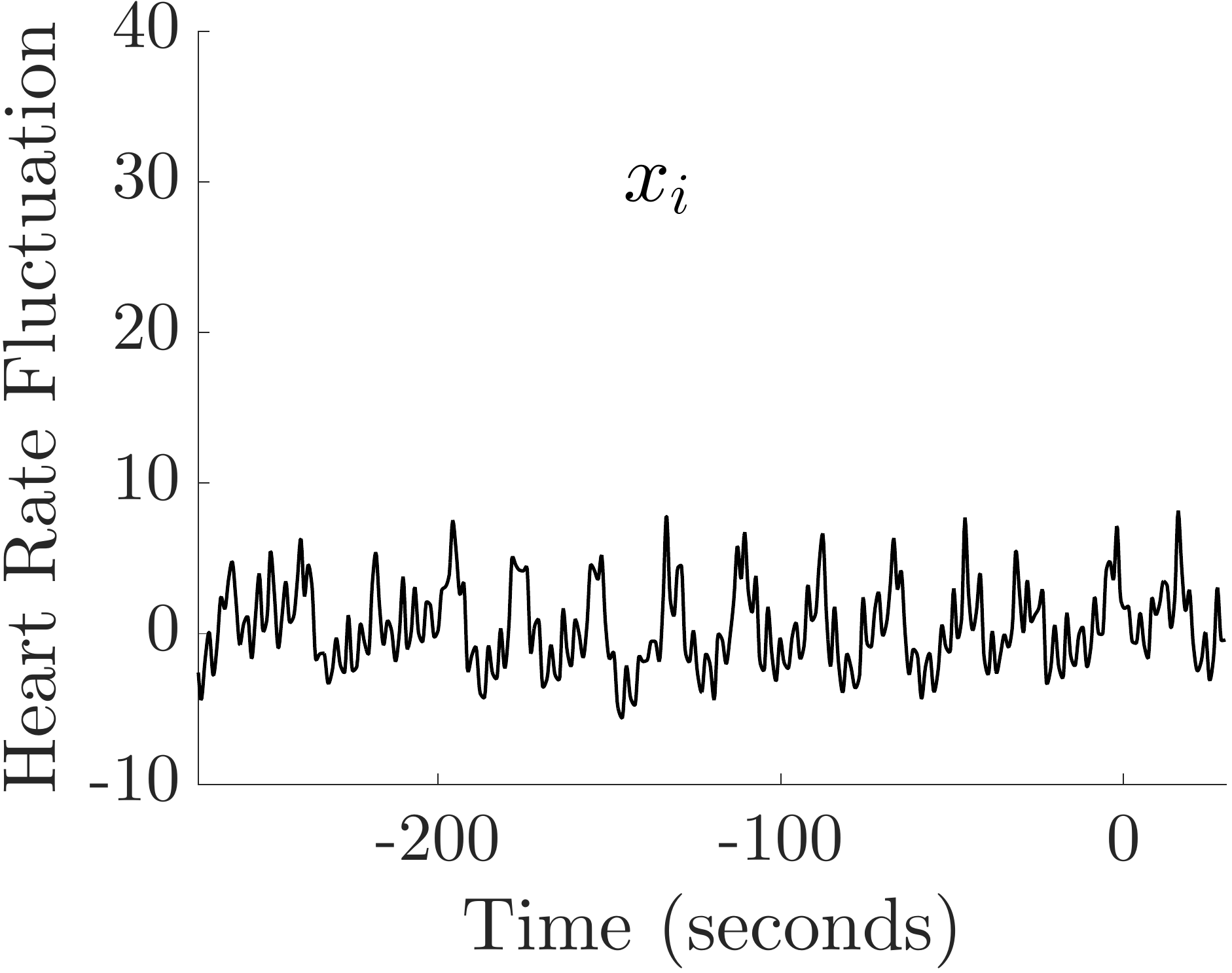}\\
\includegraphics[scale=0.2]{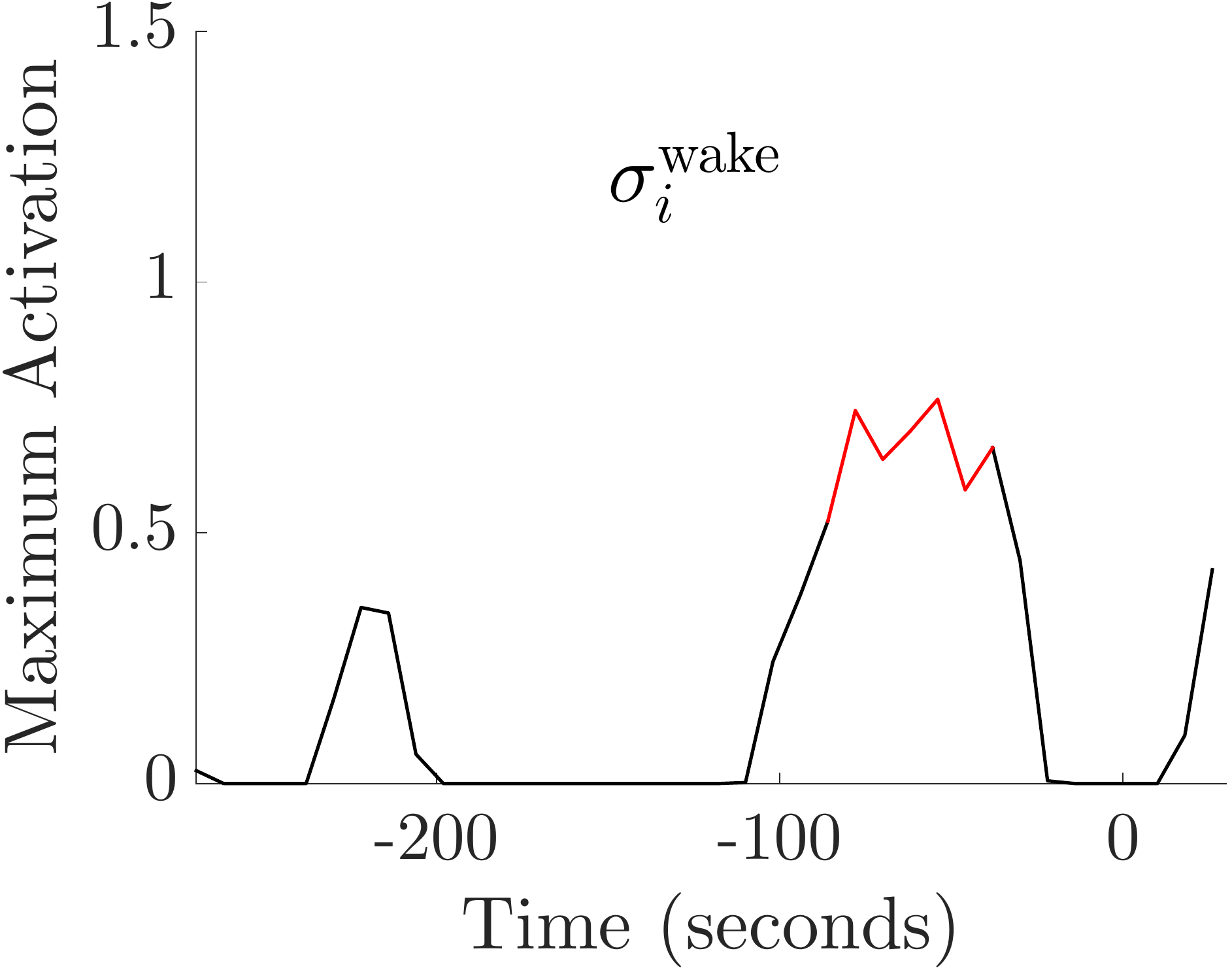}\includegraphics[scale=0.2]{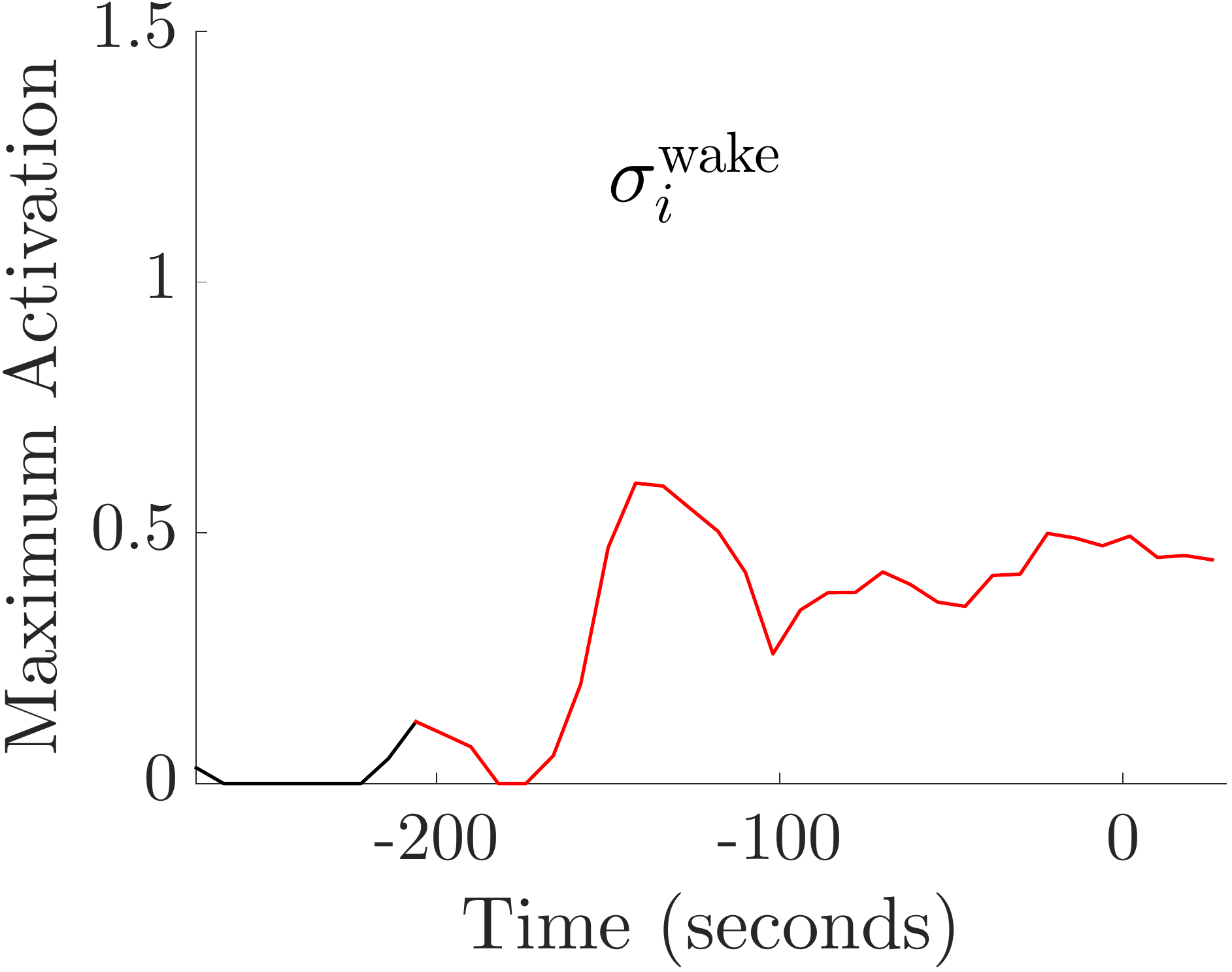}\includegraphics[scale=0.2]{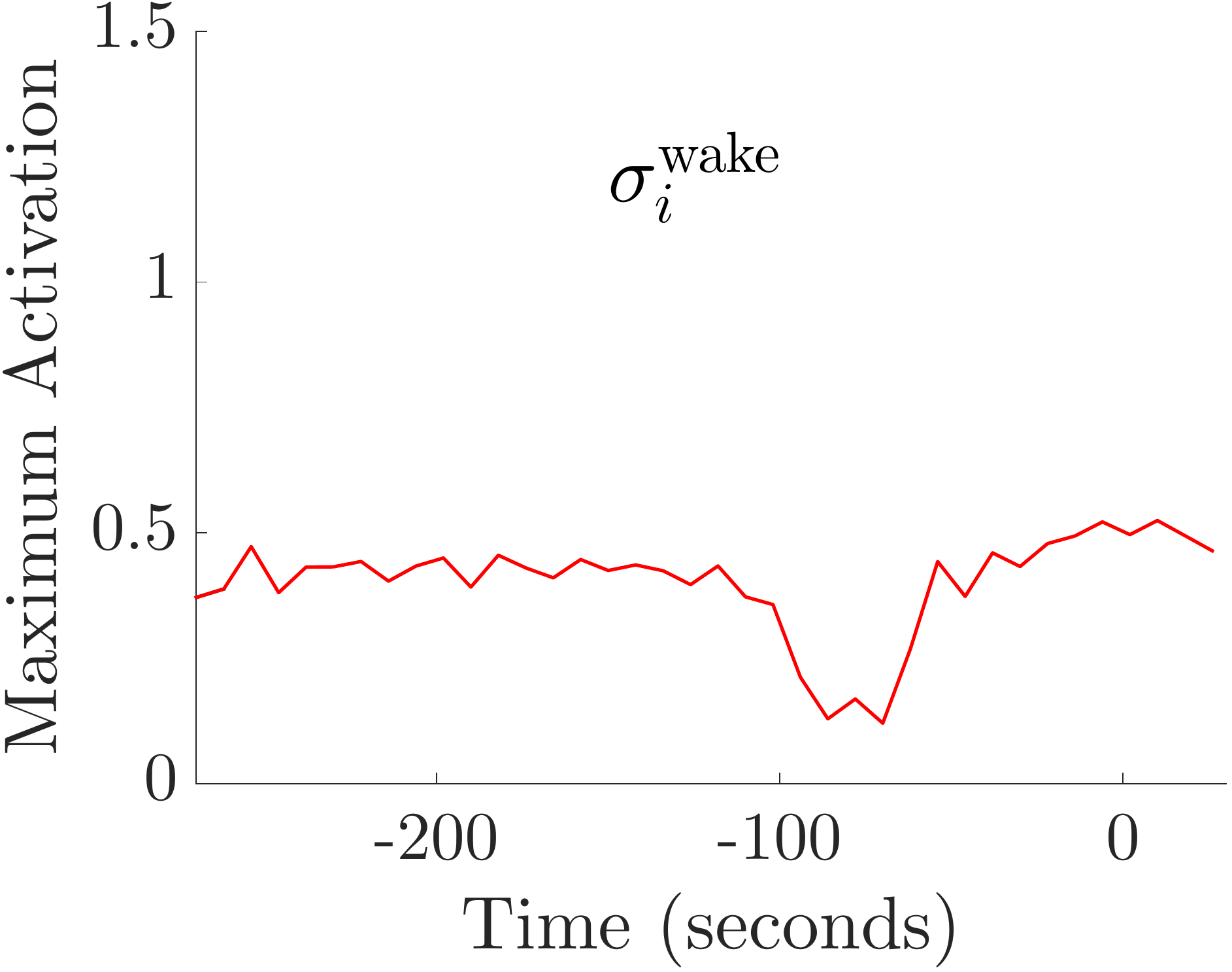}\includegraphics[scale=0.2]{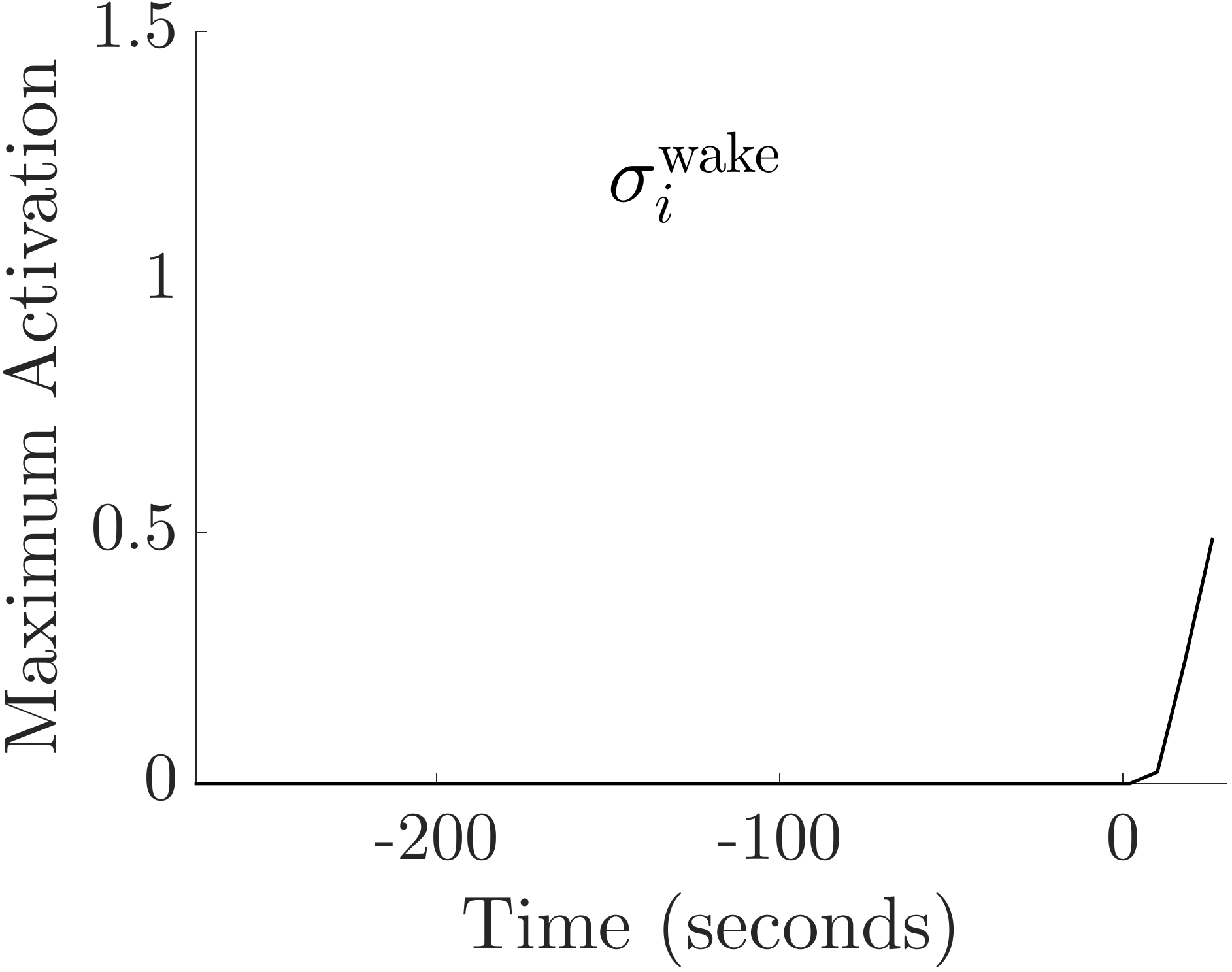}\\
\includegraphics[scale=0.2]{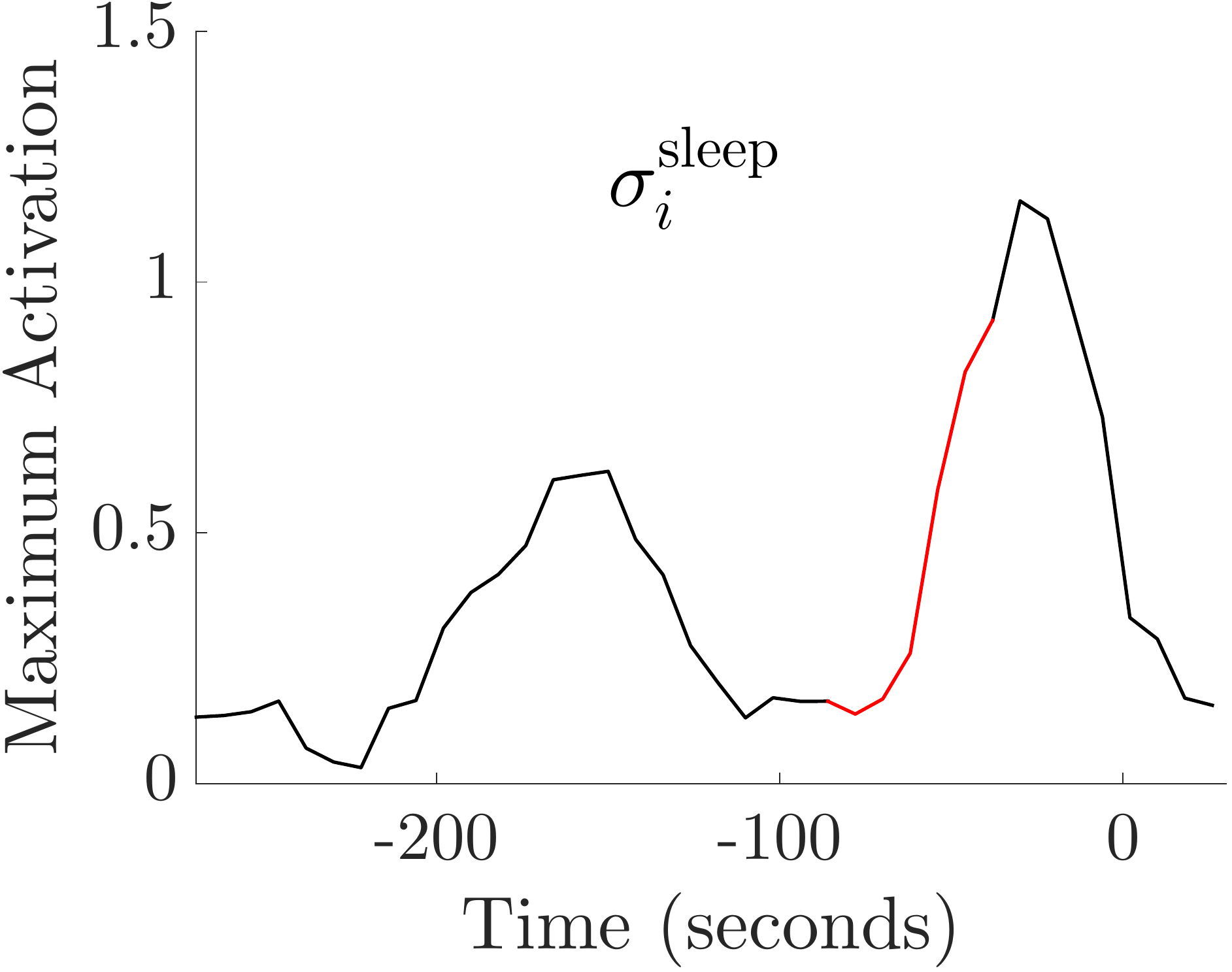}\includegraphics[scale=0.2]{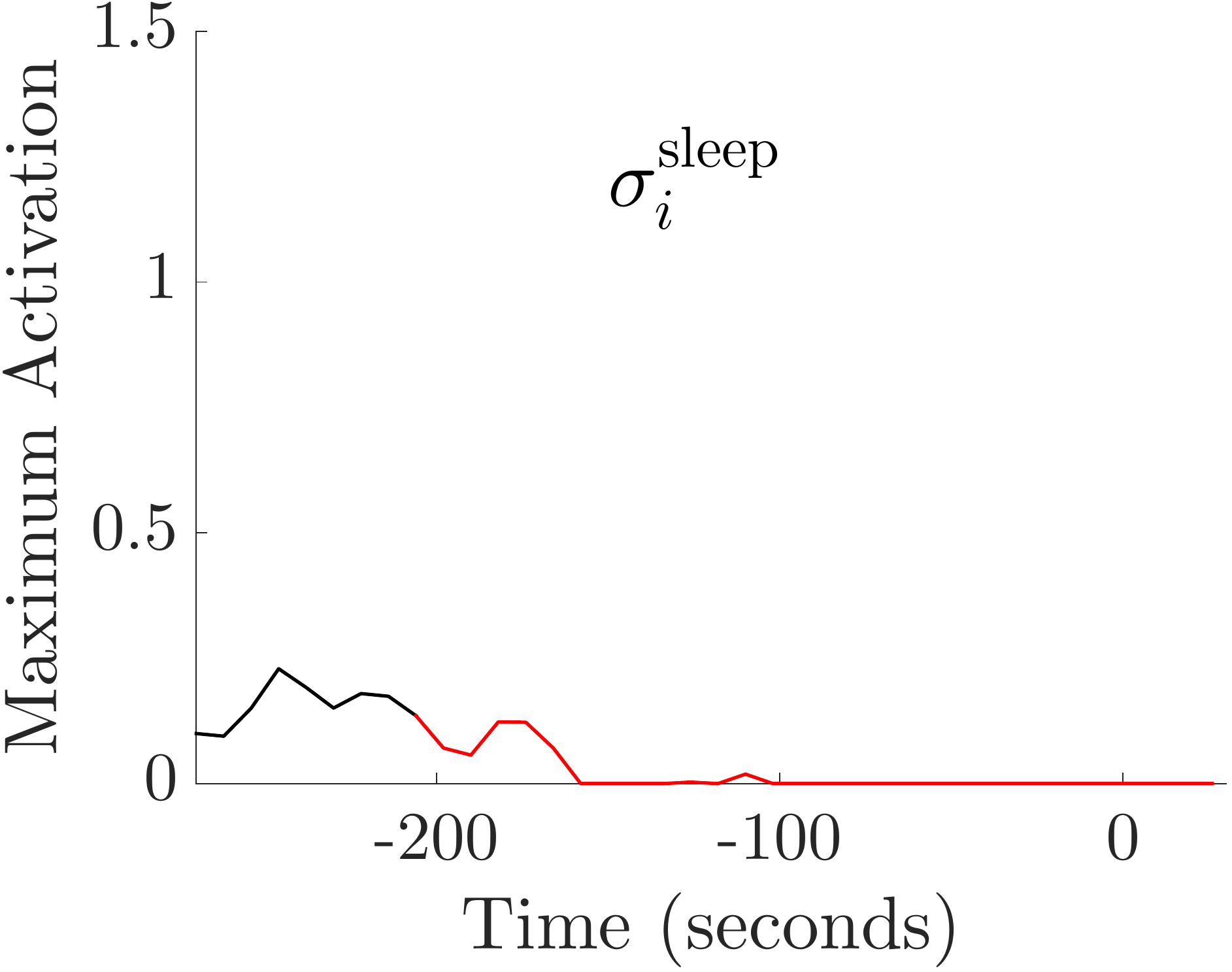}\includegraphics[scale=0.2]{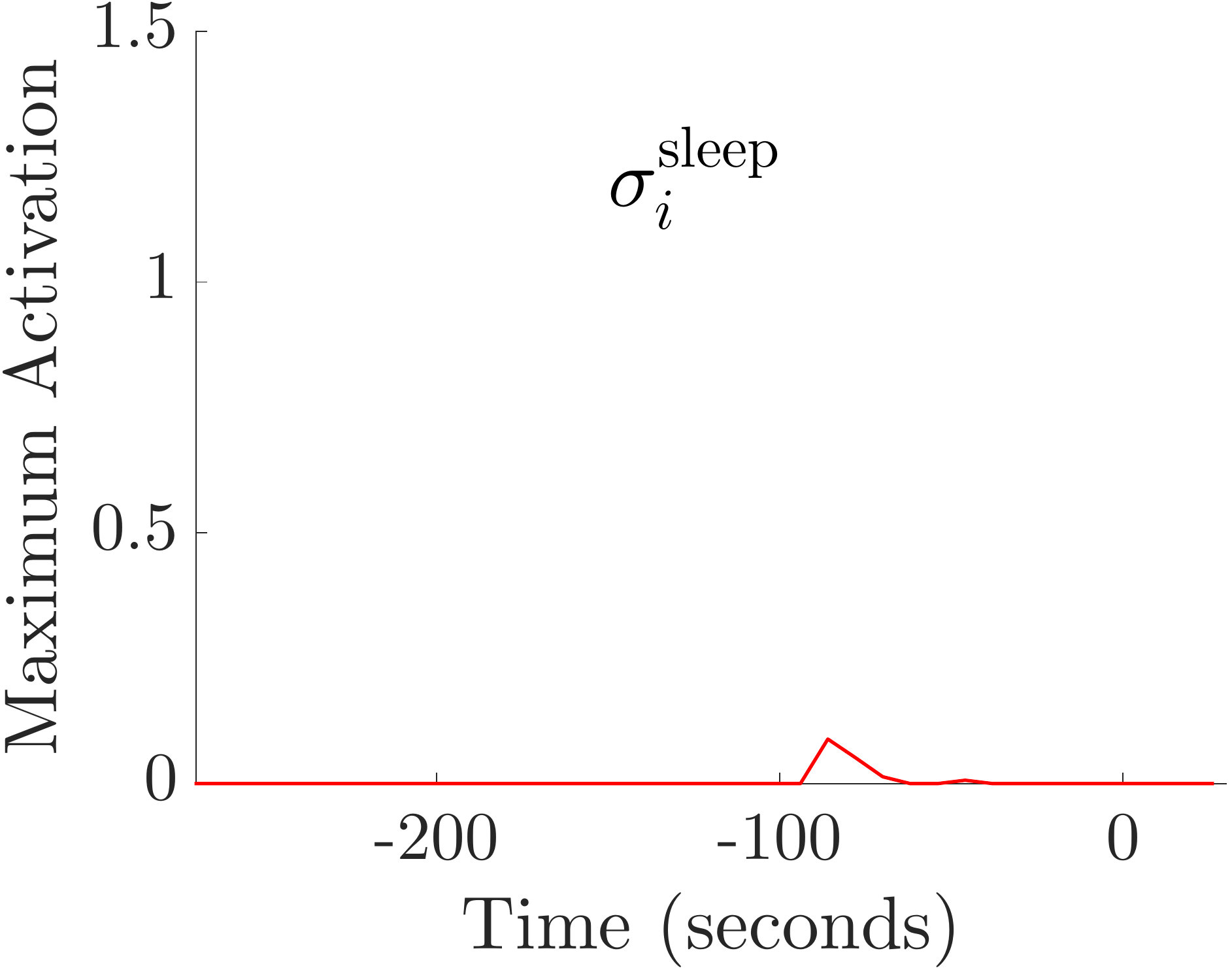}\includegraphics[scale=0.2]{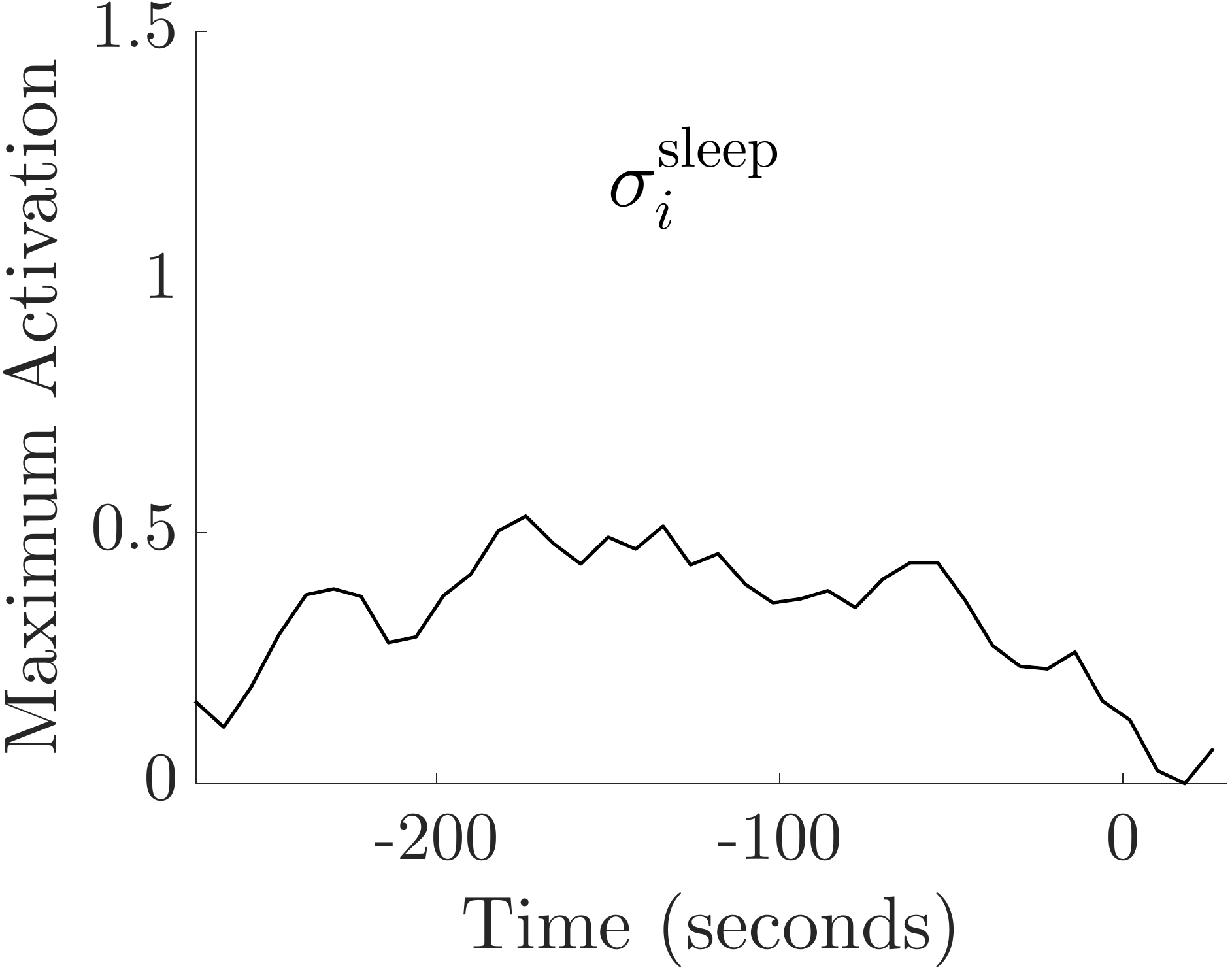}\\
\caption{\label{Fig:Activations} We present typical activations in the last convolutional layer of the CNN model. The top row shows four typical input IHR signals, and the corresponding activations in the last convolutional layer of the CNN model are shown in the bottom two rows. See \eqref{Eq:Wake} and \eqref{Eq:Sleep} for definitions of $\sigma_i^\mathrm{wake}$ and $\sigma_i^\mathrm{sleep}$. The red colour indicates that the subject is awake, whereas the black colour indicates that the subject is asleep.}
\end{figure*}

In Figure~\ref{Fig:Activations}, we feed input signals $5$ minutes in length through the trained CNN. Given an input signal $x_i$, we plot the output signals
\begin{gather}
\sigma_i^{\mathrm{wake}}(t) := \max \{ y_i^j(t) : j = 2, 3, 5, 8,9, 10 \}; \label{Eq:Wake}\\
\sigma_i^{\mathrm{sleep}}(t) := \max\{y_i^j(t) : j = 1, 4, 6, 7 \} \label{Eq:Sleep}.
\end{gather}
Note that we previously established which features $\psi_j$ in our model correspond to the wake and sleep stages (see Figure~\ref{Fig:pValue}).  Maximum activations are considered to enhance visibility.  What we see is expected: in the first three examples (from CGMH-validation), when the subject is awake, the network detects the fluctuations which, based on our physiological knowledge \cite{Snyder:1964,Somers:1993}, should correspond to sympathetic tone dominance and line up with the subject's arousal. For the first input signal, we can visualize a clear change in heart rate during the wake stage.  For the second and third signals, $\sigma_i^\mathrm{wake}$ appears to be correlated with the irregularity of the IHR series, whereas for the last signal, $\sigma_i^{\mathrm{sleep}}$ appears to be correlated with the regularity of the IHR series.  It is not always visually clear which properties of the IHR series the network is recognizing, but the results in this paper indicate that the network is successful at capturing these hidden dynamics.

\begin{figure}
\centering
\includegraphics[scale=0.3]{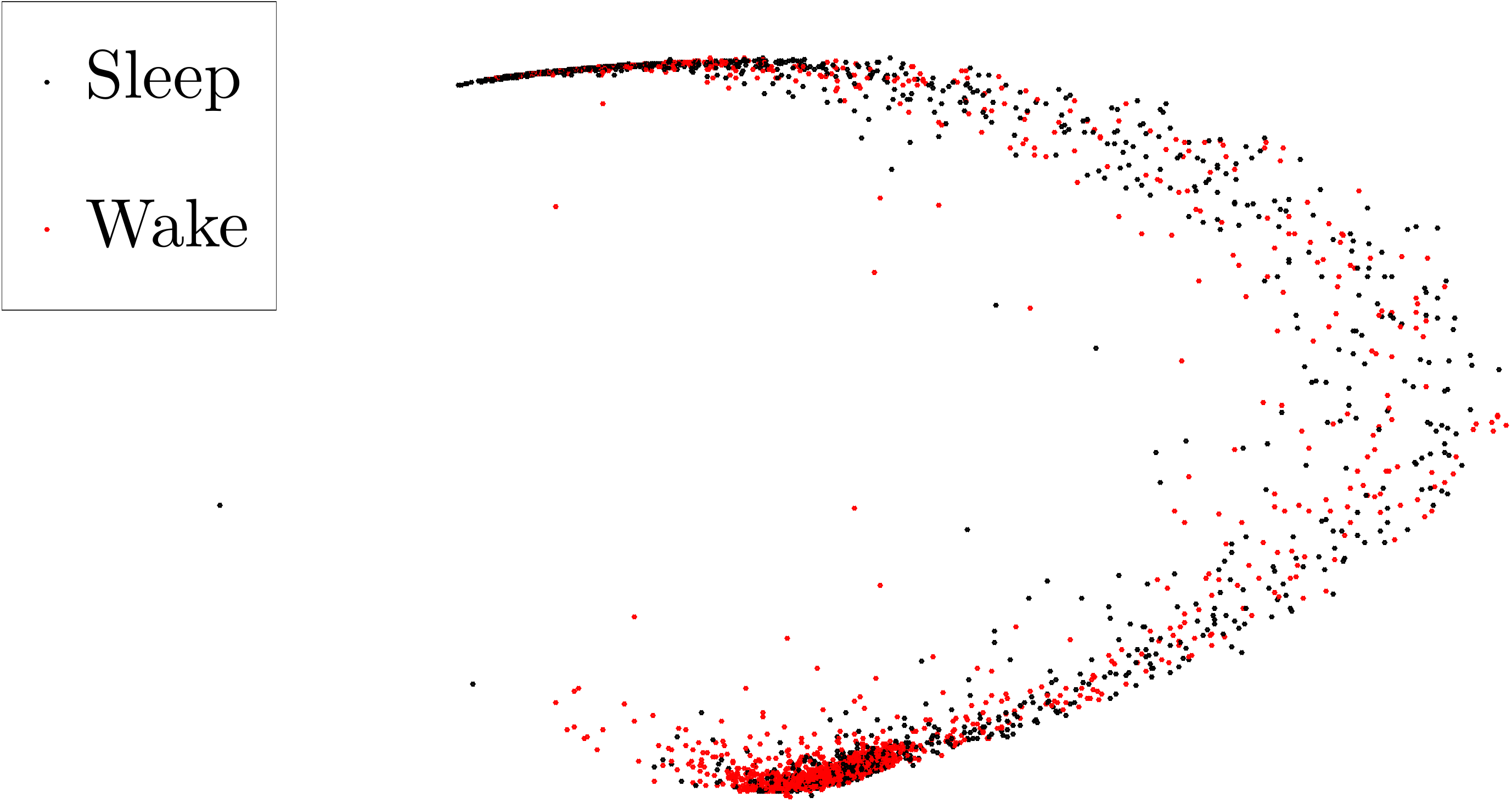}
\caption{\label{Fig:PCA}Visualization of the output of the last fully-connected layer of the CNN model by computing and projecting the data onto the first three principal components. Points are coloured according to their provided labels.}
\end{figure}

In Figure~\ref{Fig:PCA}, we visualize the performance of the CNN model using principal component analysis (PCA). Points in the plot correspond to signals from CGMH-validation, and points are close together if they have similar activations (with respect to the Euclidean norm on $\mathbb{R}^{20}$) in the last fully-connected layer of the CNN model. The proficiency of the model is seen as the plot is coloured according to label:  the positive class (red) is sufficiently separated from the negative class (black). The wake and sleep stages are clustered into two groups, with sleep stages at the top and wake stages at the bottom.  However, the model is evidently not perfect as there are several exceptions to this rule.  For visibility purposes, we only plot $40 \%$ of the black points.

\section{Discussion}\label{Section:Discussion}
In the field of HRV analysis, an HRV index is usually calculated on a window (subset) of the IHR or RRI series. For example, it is suggested to estimate HRV indices such as approximate entropy \cite{Engoren:1998} or one of the many Poincare map-based indices \cite{Casaseca-de-la-Higuera_Martin-Fernandez_Alberola-Lopez:2006} using a window containing at least $300$ beats. The choice of window size strongly affects the quality of information extracted by the chosen HRV index. If the window size is large, and the IHR can be well-approximated by a stationary process, the HRV index is stable. But in the presence of non-stationarity (as is usually the case for large windows), the index becomes unreliable (e.g. the standard deviation of a time series is strongly affected by baseline wandering). If the window size is small, under an assumption of local stationarity, the impact of non-stationarity becomes negligible. However, shortening the window destabilizes most indices because they are calculated using too little information. Hence, a type of ``uncertainty principle'' exists in HRV analysis, just as in other non-stationary time series analysis fields. 

The traditional approach to HRV index calculation is to associate a single number to each window of the IHR or RRI series. This number is effectively invariant under cyclic permutations of the window, and the location of a detected feature is not preserved. On the other hand, a $1$-dimensional convolutional layer behaves {\em equivariantly} under translations of the input, i.e. if the inputted time series is cyclically permuted, so is the outputted time series. In Figure~\ref{Fig:Activations}, we see a temporal correspondence between fluctuation morphologies in the inputted time series and large activations in the outputted time series.  The binary classifier appended to the final convolution block is informed not only of the presence of certain HRV features, but also of their location in time relative to the labeled epoch. 

The pathway through which input size affects the classification result is unclear, but we provide some interpretation. Sleep stage classification using heart rate relies on precisely locating time-varying fluctuation in the IHR series. A large-enough window is required to capture these fluctuations. 
This high-level interpretation is consistent with baseline calibration techniques used in the field of anaesthesiology. For example, in \cite{Seitsonen2005}, the authors use HRV indices to determine if a subject has felt the pain of a surgical incision during anaesthesia. Post-incision ($0\texttt{-}120 \ \mathrm{sec}$) fluctuation of the IHR series indicates a painful experience for the patient. The authors detect this fluctuation by comparing the post-incision HRV indices to the pre-incision HRV indices, while a normalization scheme is used to account for inter-individual variance. The crucial information is not the raw HRV indices but whether they begin to fluctuate after the incision compared with the baseline.  A similar study using thermal stimuli is \cite{Hamunen2012}. 

In the robustness check, we carry out a kind of ``cross-validation'' at the database level (see Table~\ref{Table:Mix}). From the statistical viewpoint, this check provides evidence that the model structure and training process is effective independently of the chosen database.  Without any theory to back up the effectiveness of the CNN framework, we need reassurance that the choices of parameters such as the number of convolution blocks, the number of filters, the kernel size, the number of epochs, and the learning rate are not over-fitted to one use-case.  We take further care to make the pre-processing steps as weak as possible: we do not manually correct the R peak detection algorithm outcome, but simply reject epochs with a very loose criterion. 
Our database-level cross-validation also shows that the selected CNN-IHR series pairing is robust to factors which should be irrelevant to the classification problem, such as the monitoring machine, the original sampling rate, and the presence of measurement noise. 
An interesting phenomenon we observe in Table~\ref{Table:Mix} is that although the UCDSADB contains subjects with sleep apnea of various severities, the model established from it performs reasonably well on CGMH-validation and the DREAMS Subjects Database.
On the other hand, when the model is trained on CGMH-training or the DREAMS Subjects Database (which are composed of normal subjects), we see that it does not perform well on the UCDSADB, particularly on those subjects with severe sleep apnea (see Figure \ref{Fig:UCDSubjects}). This result suggests an interesting conjecture: the capacity of the CNN framework is large enough to include features from both normal and abnormal subjects. In other words, the CNN efficiently learns features for both normal and abnormal subjects from the UCDSADB, but it does not learn anything about abnormal subjects from the DREAMS Subjects Database, and this conjecture explains the asymmetric validation outcomes shown in Tables \ref{Table:1Q} and \ref{Table:Mix}.

The transfer proficiency test deserves more discussion. In addition to showing that the CNN model trained on one modality can be applied to another modality, we see that the CNN model which is both trained and validated on IHR-PPG series performs slightly better than the CNN model trained and validated on IHR series (see Tables \ref{Table:1Q} and \ref{Table:PPG}).
At first glance, this result seems unrealistic because the link between the ECG signal and the ANS is more direct than the link between the PPG signal and the ANS. 
However, our least obliging intuition is that this slight difference is a consequence of the sensitivity of the PPG signal to movement. A moving subject is likely awake, and when a subject moves, his PPG signal will experience more noise than his ECG signal. Hence, on average, the IHR-PPG series will experience more distortions than the IHR series. We could hypothesize that the CNN associates such distortions with the wake stage, but other interpretations are available:  note that although the IHR and IHR-PPG series both aim to capture heart rate, they are different. The main difference comes from the {\em pulse transit time} (PTT), which describes the time it takes for pumped blood to travel from the heart to the peripheral, where the PPG signal is recorded. It is well known that the PTT contains blood pressure information and more general information about hemodynamics \cite{Mukkamala2015}. In other words, the IHR-PPG contains both heart rate information and hemodynamics information. A large scale empirical study is needed to confirm whether the CNN framework is sensitive enough to exploit this additional information, and a theoretical analysis is recommended.

Finally, we discuss our exploration of the CNN framework provided in Section~\ref{Sect:Exploration}.  The analysis carried out is applicable to only one version of the trained model.  There is no theoretical basis for expecting the model to be the same after re-training. Besides exhibiting a simple permutation of the features, the network could converge to an entirely different local minimum, causing the new feature set to be distinct from the old one. It is important to realize that although we have shown the robustness of our model from various viewpoints, this lack of explicit reproducibility experienced with NNs is troubling. 

\subsection{Comparison with Previous Work}
We compare our result to previous studies in the field that feature inter-individual validation or cross-validation.
In \cite{Xiao2013}, $41$ features were used to build a random forest model and differentiate between the wake, REM, and non-REM stages. The database was composed of healthy subjects between the ages of $16$ and $61$. 
Based on the confusion matrix provided in \cite{Xiao2013}, the sensitivity, specificity, accuracy, and F1 for detecting the wake stage are $51.2$\%, $90.2$\%, $84.0$\%, and $0.50$. However, we cannot make a direct comparison between our work and the work in \cite{Xiao2013} because the authors only analyzed those data labeled with ``stationary,'' but we analyze the whole database. 

In \cite{Mendez2010}, the REM and non-REM stages were differentiated. The authors take the temporal information into account by using a time-varying auto-regressive model, and they use the phase and magnitude of the ``sleepy pole'' as new features. The database was composed of 24 subjects between the ages of $40$ and $50$ with body mass index less than $29$ $\mathrm{kg/m}^2$ and apnea-hypopnea index $0$. The reported sensitivity, specificity, and accuracy of the trained hidden Markov model was $70.2\%$, $85.1\%$, and $79.3\%$, respectively. Although the HRV properties of the REM stage are similar to those of the wake stage, they are physiologically different and we cannot achieve a fair comparison.

In~\cite{Lewicke2008}, a variety of features and classification algorithms were considered and evaluated on a database composed of 190 infants. The wake and sleep classes were balanced for the analysis. The sensitivity and specificity of their multi-layer perceptron model (without rejection) was $79.0 \%$ and $77.5 \%$, respectively. Because there are physiological variations between the sleep dynamics of infants and adults, a comparison might not be meaningful.

In \cite{Aktaruzzaman2015}, a feed-forward NN was applied to various time-domain, frequency-domain, and regularity features to differentiate between the wake and sleep stages.  Detrended fluctuation analysis was also used. Various epoch lengths were considered, and the highest performance was recorded on an epoch length of $5$ minutes. The ECG recordings came from $20$ subjects aged $49$-$68$ years with varying degrees of sleep apnea. The authors performed two cross-validation schemes: in the inter-individual leave-one-subject-out scheme, the accuracy, sensitivity, specificity, and kappa coefficient was $71.9 \pm 18.2 \%$, $43.7 \pm 27.3 \%$, $89.0 \pm 7.8 \%$, and $0.29 \pm 0.24$, respectively. (The other validation method was not inter-individual.) In comparison with our approach, the lower performance of their model could be attributed to the physiological heterogeneity of their subjects.

In~\cite{Long2012}, an adaptive method was used to extract spectral HRV features from the IHR series.  
Fifteen subjects aged $31.0 \pm 10.4$ years with Pittsburgh Sleep Quality Index less than $6$ were considered in a leave-one-subject-out cross-validation scheme. The linear discriminant-based classifier achieved a Cohen's kappa coefficient of $0.48 \pm 0.24$ and an $\mathrm{AUC}$ of $0.54$. The sensitivity was $49.7 \pm 19.2 \%$, and the specificity was $96 \pm 3.3\%$.  
Further publications which perform inter-individual classification are \cite{Thomas2005,karlen:2009,Fonseca2015}. However, they are not compared here because they additionally report use of the respiratory signal.

Overall, due to the heterogeneity of the data sets used in different publications, it is not easy to make direct comparisons, but we recognize from the reported numbers the standard to which we should adhere.

\subsection{Limitations and Future Work}

From the database perspective, the chief limitation in this study is the class imbalance. As a result of this imbalance, the training procedure is more challenging, and the algorithm's performance is skewed by the size difference between the two classes. This can be seen in the low sensitivity of all validation checks. Another limitation is the case number. It is widely believed that the larger the high quality database we have, the better the model we can train. We show that a small training set from a few subjects is enough to obtain a network which generalizes well, but, as demonstrated by our results on the UCDSADB in Table~\ref{Table:1}, physiological heterogeneity related to sleep apnea contributes strongly to adverse classification results.  With a larger database, we might tune the training process to accommodate physiological variability among subjects and databases. 

From the signal processing perspective, in this study we focus mainly on the IHR and IHR-PPG series, which are extracted from the raw ECG and PPG signals. It is clear that the IHR and IHR-PPG series are ``dimension reduced'' versions of the raw ECG and PPG signals, which contain more information than simply heart rate; e.g. respiratory information is usually hidden in the ECG and PPG signals, and blood pressure and hemodynamics information is contained in the PPG signal. We might obtain an improvement by taking the raw time series into account.  The cost of doing so increases as both the size of the network and the size of the training set would need to grow.
On the other hand, if there are more recorded vital signs available, taking them into account may improve classification performance. For example, we might take multiple ECG leads into account, or we might consider an ECG recording and a PPG recording simultaneously. It has been shown in \cite{Thomas2005,karlen:2009,Fonseca2015} that the respiratory flow signal contains abundant information related to sleep dynamics.

From the algorithmic and theoretical perspective, the network architecture employed in this paper is relatively simple compared to other works in the field (e.g. \cite{yang2015deep}) and could be augmented to improve classification performance at the cost of computational resources. We might introduce a recurrent structure to the network, as in long short-term memory networks \cite{Hochreiter1997}, to beneficially utilize the sequential nature of the problem. Although there have been several efforts \cite{Mallat2012,Patel2016,LinRolnick2017,Wiatowski2017} to understand how CNNs and general deep NNs work, it is still a quite open question. Again, understanding this ``black box'' is an urgent mission for upcoming applications.

From the clinical perspective, accurately assessing the wake stage has an important application in the field of sleep apnea screening. The apnea-hypopnea index (AHI) is obtained by counting the apnea and hypopnea events that occur while the subject is asleep. If a patient is labeled as awake while he is asleep, then his AHI will be underestimated.  Without a proper diagnosis, an unhealthy patient will continue to be exposed to the dangers associated with severe sleep apnea.
Our established model can be applied to the problem of screening for sleep apnea using home-care class IV equipment. 
There are several open problems listed above, ranging from data collection, to theoretical work, to clinical applications. We will report the result of resolving these problems in future work.

\section{Conclusion}

In this study, we build a CNN model to predict every $30$ seconds whether a subject is awake or asleep using his overnight ECG or PPG recording. The considered CNN not only detects useful HRV features from the interpolated IHR series but also pinpoints their location in time. 
Since the DREAMS Subjects Database is composed of different races, our results suggest that the model is flexible enough to be applied to different races. 
The acceptable performance on the UCDSADB shows that the model is flexible enough to be applied to subjects with various severities of sleep apnea.
Our series of robustness checks provides evidence that although the CNN framework is minimally understood from a theoretical viewpoint, its effectiveness in our chosen application is not accidental.

%Our approach is furthermore scalable; this scalability comes from the fact that the compressed representation of the IHR series results in fewer computational resources being expended in the training process (compared to using the raw ECG or PPG). This scalability is beneficial for applications related to mobile devices. 

\bigskip

\bibliographystyle{amsplain}
\bibliography{BibSleep} 

\clearpage

\appendix
\section{Supplementary Results}

For the sake of self-containedness, we present results in connection with the main result which further validate the use of this method in the particular field of sleep stage classification.  Although it is not the main focus of this work, we have interest in the binary problem of differentiating between the REM and non-REM stages while a subject is asleep \cite{Mendez2010,Aktaruzzaman2015}. As in \cite{Aktaruzzaman2015}, we discard all epochs labeled with ``wake.''  Since we are no longer concerned with the wake stage, we return to our original database of $90$ subjects from CGMH and reject a subject if his ratio of ``REM'' to ``non-REM'' labels is less than $0.1$. There are $77$ subjects remaining, and we call this database CGMH-training (REM). (Besides removing the ``wake'' epochs, we do not change CGMH-validation for this experiment.)  We train the CNN model on signals of length $5$ minutes from CGMH-training (REM), and report in Table~\ref{Table:REM-NREM} the performance measures associated with applying the trained model to the three validation databases. In this table, the positive class is associated to the ``REM" label. After disregarding the ``wake'' labels, the percent of ``REM'' labels is $17.8\%$ in CGMH-training (REM), $14.8\%$ in CGMH-validation, $18.1\%$ in the DREAMS Subjects Database, and $19.0\%$ in the UCDSADB. Overall, the performance of the CNN model appears to exceed the performances reported in \cite{Mendez2010,Aktaruzzaman2015}. Note that the performance is lower in the UCDSADB, which is expected due to sleep disturbances caused by sleep apnea.

The scattering transform (ST) \cite{Mallat2012} is a theoretical framework based on the wavelet transform that builds representations of time series and images that are useful for classification.   The ST is inspired by the success of the CNN framework and builds features using a similar succession of operations. The crucial difference between the ST and the CNN framework is that the filters employed in the ST are pre-designed wavelets and are not learned. The theoretical properties of the ST are well-understood and are discussed in \cite{Mallat2012} with several generalizations \cite{Czaja_Li:2018,Qiu2018}. We replace the feature design step in our CNN model with the ST in order to provide evidence that the effectiveness of our CNN model is not accidental. 
We compute the second order ($m = 2$) ST of each input time series using Morlet wavelets, one wavelet per octave ($Q = 1$), and a local averaging parameter of $T = 2^7$.  A support vector machine (SVM) is commonly used to perform the classification step. However, in our work, the features designed by the ST are fed through a neural network classifier consisting of two dense layers of $20$ nodes; each node employs a bias and a sigmoid activation function. The output of the second dense layer is fed with biases into two output nodes corresponding to the wake and sleep stages. We train the network using scaled conjugate gradient back-propagation for $200$ epochs, and we use cross-entropy as the loss function. We implement all of the ST and classification steps using MATLAB R2015a, ScatNet-0.2,\footnote{\url{http://www.di.ens.fr/data/software/scatnet/}} and the Neural Network Toolbox.  After training the ST-based model on CGMH-training, the performance statistics are shown in Table~\ref{Table:ST}. The results obtained on CGMH-validation are close to the corresponding results obtained using the CNN model. The results obtained on the DREAMS Subjects Database and the UCDSADB are acceptable but not as strong as the corresponding results obtained using the CNN model.  

\begin{table}
\caption{Performance statistics for the REM vs. non-REM model trained on CGMH-training (REM)}
\label{Table:REM-NREM}
\begin{center}
\begin{small}
\begin{tabular}{lcccc}
\toprule
 & \multirowcell{3}{CGMH\\-training\\(REM)} & \multirowcell{3}{CGMH\\-validation} & \multirowcell{3}{DREAMS\\Subjects} & \multirowcell{3}{UCDSADB}\\ \\ \\
\midrule
TP   & $5,822$ & $1,563$ & $1,363$ & $1,626$ \\
FP  & $3,154$ & $1,122$ & $713$ & $1,222$ \\
TN   & $37,632$ & $13,083$ & $12,974$ & $11,527$\\
FN  & $3,017$ & $901$ & $1,656$ & $1,361$\\[0.6em]
SE & $65.9$ & $63.4$ & $45.2$ &  $54.4$\\
SP & $92.3$ & $92.1$ & $94.8$ &  $90.4$\\
ACC   & $87.6$ & $87.9$ & $85.8$ &  $83.6$ \\[0.6em]
PR  & $64.9$ & $58.2$ & $65.7$ &  $57.1$\\
F1    & $0.65$ & $0.61$ & $0.54$ &  $0.56$\\
AUC        & $0.91$ & $0.89$ & $0.87$ &  $0.83$\\
Kappa      & $0.58$ & $0.54$ & $0.45$ &  $0.47$ \\
\bottomrule
\end{tabular}
\end{small}
\end{center}
\vspace{0.1in}
\begin{footnotesize}
TP: true positive; FP: false positive; TN: true negative; FN: false negative; ACC: accuracy; AUC: area under the ROC curve; PR: precision; SE: sensitivity; SP: specificity.
\end{footnotesize}
\vspace{0.4in}
\end{table}

\begin{table}
\caption{Performance statistics for the scattering transform-based model trained on CGMH-training}
\label{Table:ST}
\begin{center}
\begin{small}
\begin{tabular}{ccccc}
\hline\\ [-0.7em]
  & \multirowcell{2}{CGMH\\-training} & \multirowcell{2}{CGMH\\-validation} & \multirowcell{2}{DREAMS} & \multirowcell{2}{UCDSADB} \\ \\ [0.2em]
\hline \\ [-0.6em]  
TP    & $4,026$ &$1,709$ & $1,569$ & $1,773$  \\
FP  & $3,874$  &$1,684$  & $2,560$ & $2,873$ \\
TN    & $29,819$ &$14,985$  & $14,123$ & $12,863$\\
FN  & $3,753$ & $1,724$ & $1,780$ & $2,465$\\[0.6em]
SE  & $51.8$ &  $49.8$ &  $46.8$ & $41.8$\\
SP & $88.5$ & $89.9$ & $84.7$ & $81.7$ \\
ACC    & $81.6$ & $83.0$ & $78.3$  & $73.3$ \\[0.6em]
PR   & $51.0$ & $50.4$ & $38.0$ & $38.2$\\
F1     & $0.51$ & $0.50$ & $0.42$ & $0.40$ \\
AUC         & $0.83$ & $0.82$ & $0.75$ & $0.68$ \\
Kappa       & $0.40$ & $0.40$ & $0.29$ & $0.23$ \\
[0.2em]
\hline
\end{tabular}
\end{small}
\end{center}
\vspace{0.1in}
\begin{footnotesize}
TP: true positive; FP: false positive; TN: true negative; FN: false negative; ACC: accuracy; AUC: area under the ROC curve; PR: precision; SE: sensitivity; SP: specificity.
\end{footnotesize}
\end{table}

\end{document}